\documentclass{emulateapj}
\usepackage{apjfonts}
\journalinfo{The Astrophysical Journal, 679, 2008 May 20, in press}

\shorttitle{PREFERENTIAL ION HEATING IN THE CORONA}
\shortauthors{CRANMER, PANASYUK, \& KOHL}

\begin{document}

\title{Improved Constraints on the Preferential Heating and
Acceleration of Oxygen Ions in the Extended Solar Corona}

\author{Steven R. Cranmer, Alexander V. Panasyuk,
and John L. Kohl}
\affil{Harvard-Smithsonian Center for Astrophysics,
60 Garden Street, Cambridge, MA 02138 \\
Submitted 2007 November 27; \, accepted 2008 January 29}

\begin{abstract}
We present a detailed analysis of oxygen ion velocity distributions
in the extended solar corona, based on observations made with the
Ultraviolet Coronagraph Spectrometer (UVCS) on the {\em SOHO}
spacecraft.
Polar coronal holes at solar minimum are known to exhibit
broad line widths and unusual intensity ratios of the
\ion{O}{6} $\lambda\lambda$1032, 1037 emission line doublet.
The traditional interpretation of these features has been that
oxygen ions have a strong temperature anisotropy, with the
temperature perpendicular to the magnetic field being much larger
than the temperature parallel to the field.
However, recent work by Raouafi and Solanki suggested that it may
be possible to model the observations using an isotropic velocity
distribution.
In this paper we analyze an expanded data set to show that the
original interpretation of an anisotropic distribution is the
only one that is fully consistent with the observations.
It is necessary to search the full range of ion plasma parameters
to determine the values with the highest probability of agreement
with the UVCS data.
The derived ion outflow speeds and perpendicular kinetic
temperatures are consistent with earlier results, and there
continues to be strong evidence for preferential ion heating and
acceleration with respect to hydrogen.
At heliocentric heights above 2.1 solar radii, every UVCS data
point is more consistent with an anisotropic distribution
than with an isotropic distribution.
At heights above 3 solar radii, the exact probability of
isotropy depends on the electron density chosen to simulate the
line-of-sight distribution of \ion{O}{6} emissivity.
The most realistic electron densities (which decrease steeply
from 3 to 6 solar radii) produce the lowest probabilities of
isotropy and most-probable temperature anisotropy ratios that
exceed 10.
We also use UVCS \ion{O}{6} absolute intensities to compute the
frozen-in O$^{5+}$ ion concentration in the extended corona;
the resulting range of values is roughly consistent with recent
downward revisions in the oxygen abundance.
\end{abstract}


\section{Introduction}

The physical processes that heat the solar corona and accelerate
the solar wind are not yet understood completely.
In order to construct and test theoretical models, there must
exist accurate measurements of relevant plasma parameters in
the regions that are being heated and accelerated.
In the low-density, open-field regions that reach into
interplanetary space, the number of plasma parameters that need
to be measured increases because the plasma begins to become
collisionless and individual particle species (e.g., protons,
electrons, and heavy ions) can exhibit different properties.
Such differences in particle velocity distributions are valuable
probes of ``microscopic'' processes of heating and acceleration.
The Ultraviolet Coronagraph Spectrometer (UVCS) operating aboard
the {\em Solar and Heliospheric Observatory} ({\em{SOHO}})
spacecraft has measured these properties for a variety of
open-field regions in the extended corona
(Kohl et al.\  1995, 1997, 2006).

In this paper we focus on UVCS observations of heavy ion emission
lines (specifically \ion{O}{6} $\lambda\lambda$1032, 1037) in
polar coronal holes at solar minimum.
One main goal is to resolve a recent question that has arisen
regarding the existence of anisotropic ion temperatures
in polar coronal holes.
Several prior analyses of UVCS data have concluded that there
must be both intense preferential heating of the O$^{5+}$ ions,
in comparison to hydrogen, and a strong field-aligned anisotropy
with a much larger temperature in the direction perpendicular to
the magnetic field than in the parallel direction (see,
e.g., Kohl et al.\  1997, 1998; Li et al.\  1998;
Cranmer et al.\  1999; Antonucci et al.\  2000;
Zangrilli et al.\  2002; Antonucci 2006; Telloni et al.\  2007).
However, Raouafi \& Solanki (2004, 2006) and Raouafi et al.\  (2007)
have reported that there may not be a compelling need for
O$^{5+}$ anisotropy depending on the assumptions made about the
other plasma properties of the coronal hole (e.g., electron
density).

The determination of O$^{5+}$ preferential heating, preferential
acceleration, and temperature anisotropy has spurred a great deal
of theoretical work (see reviews by Hollweg \& Isenberg 2002;
Cranmer 2002a; Marsch 2005; Kohl et al.\  2006).
It is thus important to resolve the question of whether these
plasma properties are definitively present on the basis of the
UVCS/{\em{SOHO}} observations.
In this paper, we attempt to analyze all possible combinations
of O$^{5+}$ properties (number density, outflow speed, parallel
temperature, and perpendicular temperature) with the full
effects of the extended line of sight (LOS) taken into account.
The applicability of any particular combination of ion properties
is evaluated by computing a quantitative probability of agreement
between the modeled set of emission lines and a given observation.
Preliminary results from this work were presented by
Cranmer et al.\  (2005) and Kohl et al.\  (2006).

The original UVCS results of preferential ion heating and
acceleration---as well as strong ion temperature anisotropy
($T_{\perp} \gg T_{\parallel}$)---were somewhat surprising,
but these extreme departures from thermal equilibrium are
qualitatively similar to conditions that have been measured for
decades in high-speed streams in the heliosphere.
At their closest approaches to the Sun ($\sim$ 0.3 AU), the
{\em Helios} probes measured substantial proton temperature
anisotropies with $T_{\perp} > T_{\parallel}$
(Marsch et al.\  1982; Feldman \& Marsch 1997).
In the fast wind, most ion species also appear to flow faster
than the protons by about an Alfv\'{e}n speed ($V_A$), and this
velocity difference decreases with increasing radius and
decreasing proton flow velocity (e.g., Hefti et al.\  1998;
Reisenfeld et al.\  2001).
The temperatures of heavy ions are significantly larger than
proton and electron core temperatures.
In the highest-speed wind streams, ion temperatures exceed
simple mass proportionality with protons (i.e., heavier ions
have larger most-probable speeds), with
$(T_{\rm ion} / T_{p}) > (m_{\rm ion} / m_{p})$,
for $m_{\rm ion} > m_{p}$ (e.g., Collier et al.\  1996).
UVCS provided the first evidence that these plasma properties
are already present near the Sun.

The outline of this paper is as follows.
In {\S}~2 we present an expanded collection of UVCS/{\em{SOHO}}
observational data that is used to determine the O$^{5+}$
ion properties.
{\S}~3 outlines the procedure we have developed to produce
empirical models of the plasma conditions in polar coronal holes
and to compute the probability of agreement between any given
set of ion properties and the observations.
The resulting ranges of ion properties that are consistent with
the UVCS observations are presented in {\S}~4 along with, in our
view, a resolution of the controversy regarding the oxygen
temperature anisotropy.
Finally, {\S}~5 gives a summary of the major results of this
paper and a discussion of the implications these results may
have on theoretical models of coronal heating and solar wind
acceleration.

\section{Observations}

The UVCS instrument contains three reflecting
telescopes that feed two ultraviolet toric-grating spectrometers
and one visible light polarimeter (Kohl et al.\  1995, 1997).
Light from the bright solar disk is blocked by external and
internal occulters that have the same linear geometry as the
spectrometer slits.
The slits are oriented in the direction tangent to the solar limb.
They can be positioned in heliocentric radius $r$ anywhere between
about 1.4 and 10 solar radii ($R_{\odot}$) and rotated around the
Sun in position angle.
The slit length projected on the sky is {40\arcmin}, or
approximately 2.5 $R_{\odot}$ in the corona, and the slit width
can be adjusted to optimize the desired spectral resolution
and count rate.

The UVCS data discussed in this paper consist of a large ensemble
of observations of polar coronal holes from the last solar
minimum (1996--1997).
The solar magnetic field is observed to exist in a nearly
axisymmetric configuration at solar minimum, with open field
lines emerging from the north and south polar regions and
expanding superradially to fill a large fraction of the
heliospheric volume.
The plasma properties in polar coronal holes remain reasonably
constant in the year or two around solar minimum (see, e.g.,
Kohl et al.\  2006), so we assemble the data over this time
into a single function of radius.
In this paper, we limit ourselves to the analysis of observations
of the \ion{O}{6} $\lambda\lambda$1032, 1037 emission line
doublet in these polar regions.
The relevant UVCS observations are taken from the following
three sources.
\begin{enumerate}
\item
The empirical model study of Kohl et al.\  (1998) and
Cranmer et al.\  (1999) covered the period between 1996 November
and 1997 April and took the north and south polar coronal
hole properties to be similar enough to treat them together.
\item
A detailed analysis of the north polar coronal hole by
Antonucci et al.\  (2000) coincided with the second {\em SOHO}
joint observing program (JOP 2) on 1996 May 21.
\item
We searched the UVCS/{\em{SOHO}} archive for any other
north or south polar hole observations having sufficient
count-rate statistics to be able to measure the \ion{O}{6}
line widths at radii above 2 $R_{\odot}$.
A total of 14 new or reanalyzed data points were identified
between 1996 June and 1997 July.
\end{enumerate}
The remainder of this section describes the data reduction for
the third group of new data points.
Table 1 provides details of these 14 measurements.

\begin{deluxetable*}{lccccccc}
\tablecaption{Newly Analyzed Polar Coronal Hole Data: 1996--1997}
\tablewidth{0pt}

\tablehead{
\colhead{Start Date, UT Time} &
\colhead{Obs.\  Height} &
\colhead{Slit Length\tablenotemark{a}} &
\colhead{Slit Width} &
\colhead{Exposure} &
\colhead{$V_{1/e}$} &
\colhead{${\cal R}$} &
\colhead{$I_{\rm tot}$, 1032 {\AA} line} \\
\colhead{} &
\colhead{($R_{\odot}$)} &
\colhead{(arcmin)} &
\colhead{($\mu$m)} &
\colhead{Time (hr)} &
\colhead{(km s$^{-1}$)} &
\colhead{} &
\colhead{(10$^6$ phot s$^{-1}$ cm$^{-2}$ sr$^{-1}$)}
}

\startdata

1996 Jun 21, 16:55 & 2.07 & 15.8 (N) & 75 & 9.1 &
 $363\pm 21.7$ & $2.09 \pm 0.4 $ & $32.2 \pm 6.8 $ \\
1996 Sep 29, 15:26 & 2.08 & 17.5 (N) & 75 & 5.6 &
 $473\pm 12.4$ & $1.87 \pm 0.14$ & $43.2 \pm 8.7 $ \\
1996 Nov 07, 23:38 & 2.07 & 29.7 (N) & 75 & 7.6 &
 $409\pm 18.8$ & $1.78 \pm 0.2 $ & $44.6 \pm 9.1 $ \\
1996 Nov 10, 06:30 & 3.00 & 25.7 (N) & 350 & 9.4 &
 $475\pm 27.0$ & $1.16 \pm 0.18$ & $2.08 \pm 0.44$ \\
1996 Nov 10, 15:55 & 2.56 & 25.9 (N) & 350 & 10.8 &
 $505\pm 30.5$ & $1.17 \pm 0.17$ & $4.83 \pm 1.0 $ \\
1996 Nov 16, 16:56 & 2.17 & 29.6 (N) & 340 & 9.5 &
 $417\pm 7.43$ & $1.485\pm 0.06$ & $25.1 \pm 5.0 $ \\
1997 Jan 05, 21:00 & 3.07 & 28.0 (N) & 100 & 17.6 &
 $690\pm 87.2$ & $0.957\pm 0.45$ & $2.02 \pm 0.63$ \\
1997 Jan 10, 14:54 & 3.08 & 20.8 (N) & 150 & 68.8 &
 $686\pm 38.7$ & $1.23 \pm 0.3 $ & $3.37 \pm 0.76$ \\
1997 Jan 24, 16:03 & 3.08 & 20.6 (N) & 150 & 70.8 &
 $594\pm 41.6$ & $1.01 \pm 0.3 $ & $1.66 \pm 0.40$ \\
1997 Mar 09, 18:00 & 2.57 & 19.0 (S) & 300 & 8.9 &
 $500\pm 22.0$ & $1.15 \pm 0.12$ & $5.66 \pm 1.2 $ \\
1997 Jun 04, 16:49 & 2.56 & 18.8 (N) & 300 & 9.0 &
 $527\pm 33.9$ & $0.958\pm 0.15$ & $4.14 \pm 0.90$ \\
1997 Jun 08, 20:15 & 3.10 & 18.7 (N) & 300 & 8.3 &
 $645\pm 65.6$ & $1.33 \pm 0.44$ & $1.33 \pm 0.33$ \\
1997 Jul 01, 15:50 & 2.56 & 17.2 (N) & 342 & 9.8 &
 $534\pm 56.8$ & $0.988\pm 0.26$ & $3.26 \pm 0.80$ \\
1997 Jul 04, 16:45 & 3.63 & 18.9 (N) & 342 & 29.5 &
 $451\pm 47.9$ & $1.49 \pm 0.43$ & $0.559\pm 0.13$ \\
\enddata
\tablenotetext{a}{Data were integrated over the specified slit
length.  The slit was oriented tangent to either the north (N) or
the south (S) heliographic pole, as indicated.}

\end{deluxetable*}

The criteria for identifying new UVCS data were as follows.
We adopted a time period from 1996 April, the beginning of
primary science operations, to 1998 January, after which
the new cycle's activity began to rise and high-latitude
streamers appeared regularly to signal the end of true
solar minimum.
Only measurements of the \ion{O}{6} lines above the poles
(i.e., position angles within {$\pm$15\arcdeg} of the north or
south poles) and at heights above 2 $R_{\odot}$ were
sought.\footnote{Below $r \approx 2 \, R_{\odot}$, the existing
data appear to be adequate, uncertainties are low, and there is
not much of an intrinsic spread in the intensities and line
widths as a function of height.
Also, earlier analyses did not show that collisionless effects
(ion temperature anisotropies, preferential ion heating, or
differential flow) became strong until above this height.}
Prior experience with the count rates at large heights in coronal
holes refined the search further to use measurements only with
relatively long exposure times (see Table 1) to gather
sufficient statistics to measure the line widths.
There were two observations that appeared initially to
satisfy the above criteria (1997 April 15, at 4.14 $R_{\odot}$,
and 1997 July 2, at 3.10 $R_{\odot}$), but they were not used
because the count rate statistics were inadequate for reliable
line widths.
The only point of overlap between the data in Table 1 and prior
analyses (e.g., Cranmer et al.\  1999) concerns the end of the
month-long study of the north polar coronal hole in
1997 January (at $r \approx 3 \, R_{\odot}$).
These data were reanalyzed with a different line fitting
technique and an improved UVCS pointing correction;
the computed line widths are similar to those presented by
Cranmer et al.\  (1999) and the intensities are given here
for the first time.
For completeness, though, both the old and new data points are
kept in the full ensemble of \ion{O}{6} data used below.

To achieve the lowest uncertainties in the determinations of
the \ion{O}{6} $\lambda\lambda$1032, 1037 intensities and line
widths, we typically integrated over {15\arcmin} to {30\arcmin}
along the slit (see Table 1).
This corresponds to $\pm 0.5$--1 $R_{\odot}$ on either side
of the north-south axis.
The use of such large areas implies that narrow flux-tube
structures such as dense {\em polar plumes} and the less-dense
interplume regions were not resolved.
As long as all steps of the analysis remain consistent with
such a coarsely averaged state (e.g., the use of a similarly
averaged electron density), though, this need not be a problem.
The derived plasma properties thus describe the average
conditions inside coronal holes at solar minimum and do not
address differences between plumes and interplume regions.

Details concerning the analysis of UVCS data are given by
Gardner et al.\  (1996, 2000, 2002) and Kohl et al.\  (1997,
1999, 2006).
The UVCS Data Analysis Software (DAS) was used to remove image
distortion and to calibrate the data in wavelength and intensity.
The coronal line profiles are broadened by various instrumental
effects.
The optical point spread function of the spectrometer depends on
the slit width used (with 270.3 $\mu$m corresponding to 1 {\AA}
in the spectrum), the on-board data binning, the exposed
mirror area, and the intrinsic quantization error of the
detector.
This broadening is taken into account by adjusting the line
widths of Gaussian fits to the coronal components of the data;
the data points themselves are not corrected.
Tests have shown that the coronal line width can be recovered
accurately even when the total instrumental width is within
about a factor of two of the width of the coronal component.
Instrument-scattered stray light from the solar disk is modeled
as an additional narrow Gaussian component with an intensity
and profile shape constrained by the known stray light
properties of the instrument.

The analysis of the \ion{O}{6} emission line doublet involves
four basic observable quantities:
the total intensities of the two lines and their $1/e$
Gaussian half-widths $\Delta\lambda_{1/e}$.
The latter quantities are typically expressed in Doppler
velocity units as $V_{1/e} = c \Delta\lambda_{1/e} / \lambda_{0}$,
where $\lambda_{0}$ is the rest wavelength of the line and
$c$ is the speed of light.
Rather than give the two total intensities, Table 1 provides
the total intensity $I_{\rm tot}$ of the \ion{O}{6} $\lambda$1032
line and the ratio ${\cal R}$ of the $\lambda$1032 to the
$\lambda$1037 intensities.
The uncertainties given in Table 1 take account of both
Poisson count-rate statistics and the fact that the various
instrumental corrections are known only to finite levels of
precision.
Note that the ratio ${\cal R}$ does not depend on the absolute
intensity calibration of the instrument.

UVCS/{\em{SOHO}} has not been able to resolve any departures from
Gaussian shapes for the \ion{O}{6} lines in large polar coronal holes,
so the profiles are described by just the one parameter $V_{1/e}$.
For the measurements given in Table 1, we performed the line
fitting by constraining the coronal components of the
$\lambda$1032 and $\lambda$1037 lines to have the same width.
Thus, the $V_{1/e}$ values given in Table 1 are formally a
weighted mean between the two components.
This is done mainly to lower the statistical uncertainties
but there is some observational justification for assuming
that the two components have the same width.
In situations where the count rates are high, it is difficult
to see any significant or systematic difference between the
line widths of the two components.
There are various reasons why they may be different from one
another in some regions (e.g., Cranmer 2001; Morgan \& Habbal 2004),
but more work needs to be done to identify such subtle effects.

\begin{figure}
\epsscale{1.13}
\plotone{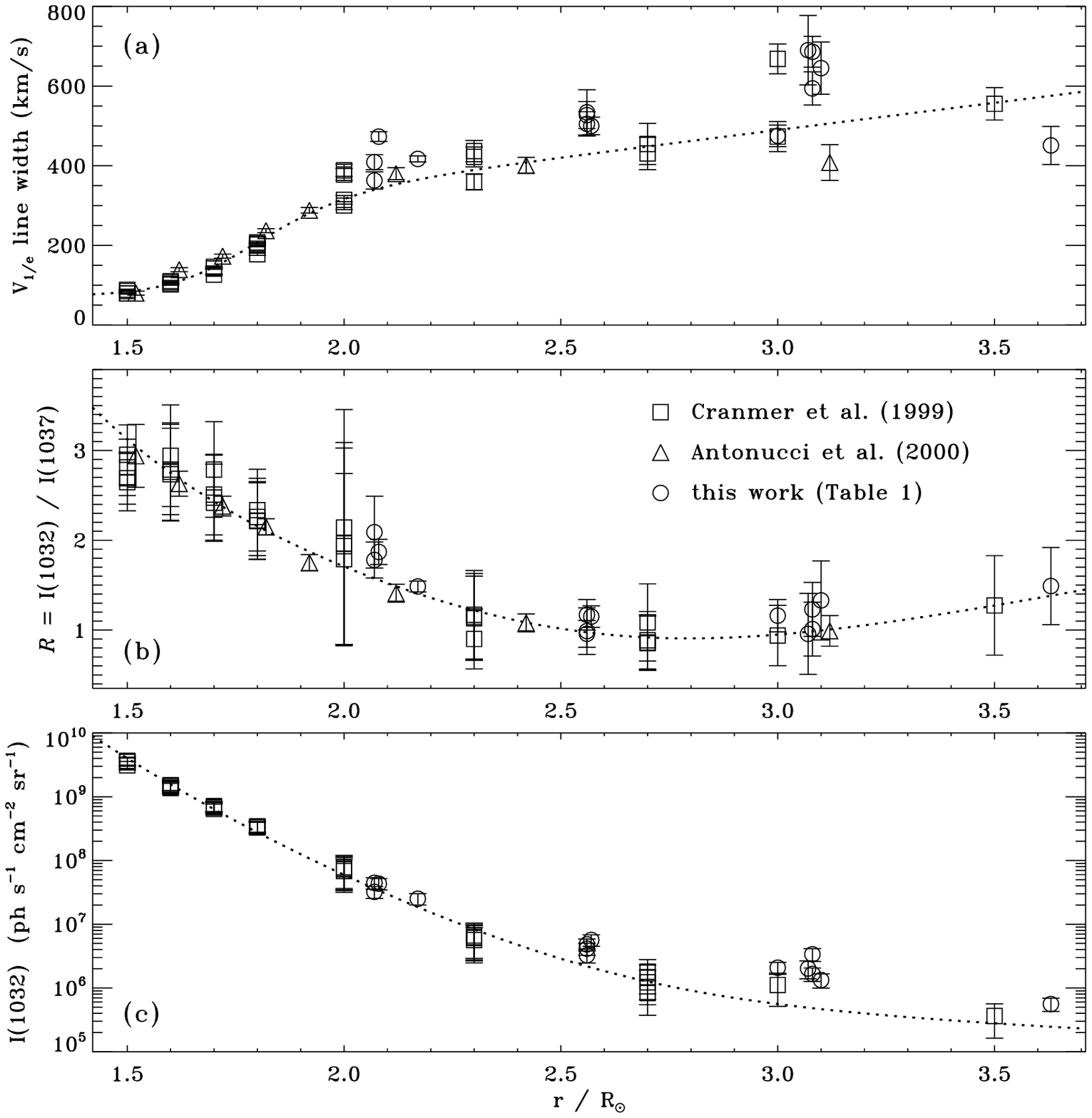}
\caption{Collected UVCS polar coronal hole measurements of
{\em (a)} \ion{O}{6} line widths $V_{1/e}$,
{\em (b)} ratio of \ion{O}{6} $\lambda$1032 to
\ion{O}{6} $\lambda$1037 intensities, and
{\em (c)} \ion{O}{6} $\lambda$1032 line-integrated intensities,
with symbols specifying the sources of the data (see labels
for references).
Error bars denote $\pm 1 \sigma$ observational uncertainties.
Also shown ({\em dotted lines}) are the parameterized fits
given by Cranmer et al.\  (1999).
{\bf SEE LAST PAGE OF PAPER FOR LARGER VERSION.}}
\end{figure}

Figure 1 displays the combined ensemble of old and new UVCS
\ion{O}{6} data for the three main observables:
the line width $V_{1/e}$, the dimensionless intensity ratio
${\cal R}$, and the absolute (line-integrated) intensity of
the \ion{O}{6} $\lambda$1032 line.
There are a total of 53 separate data points from the three
sources discussed above, but not all of these points have all
three of the main quantities: there are 50 values of $V_{1/e}$,
52 values of ${\cal R}$ (with only 49 cases where both
$V_{1/e}$ and ${\cal R}$ exist for the same measurement),
and 44 values of $I_{\rm tot}$.
This relative paucity of data illustrates the difficulties of
measuring the plasma parameters at large heights in polar
coronal holes.

In general, the radial dependences of the \ion{O}{6} quantities
in Figure 1 are similar to those given by Kohl et al.\  (1998),
Cranmer et al.\  (1999), and Antonucci et al.\  (2000).
There exists a reasonably large spread in the $V_{1/e}$ values
in Figure 1{\em{a}} above $r \approx 2.5 \, R_{\odot}$.
This spread exceeds the magnitude of the $\pm 1 \sigma$
uncertainty limits for the individual measurements, and
thus seems to indicate that there is an {\em intrinsic
variability} (possibly temporal) of the O$^{5+}$ plasma
conditions in polar coronal holes above heights where the
ions become collisionless.
It is possible that polar plumes and interplume regions
become collisionless over different ranges of radius, and thus
preferential ion heating mechanisms may begin to broaden the
\ion{O}{6} lines at different rates in the two regions.
The observed variation in line width may thus depend
on the relative concentrations of plume and interplume plasma
along the line of sight at different observation times.

\section{Empirical Model Procedure}

The observable properties of the \ion{O}{6} line doublet depend
on a nontrivial combination of various O$^{5+}$ plasma
parameters, as well as electron parameters, integrated along the
optically thin line of sight.
In general, then, it is not possible to derive accurate and
self-consistent plasma parameters via a simple ``inversion''
from the line widths and intensities.
Rather, one must build up a so-called {\em empirical model} of
the coronal hole---with the O$^{5+}$ velocity distribution and
other properties as free parameters---and synthesize trial
line profiles.
After some procedure of varying the coronal parameters to
achieve agreement between the synthesized line profiles and the
observations, the self-consistent empirical model of the ion
properties can be considered complete.
This technique is closely related to forward modeling
approaches being used in other areas of solar physics
(e.g., Judge \& McIntosh 1999).

The use of the term ``empirical model'' has resulted in a bit of
confusion regarding what assumptions are embedded in the
derived plasma parameters.
We emphasize that the empirical models described here do not
specify the physical processes that maintain the coronal plasma
in its assumed steady state.
Thus, there is no explicit dependence on ``theoretical'' concepts
such as coronal heating and acceleration mechanisms, waves and
turbulent motions, or magnetohydrodynamics (MHD), within the
empirical models.
The derived O$^{5+}$ plasma parameters depend on only the
observations and on well-established theory such as the radiative
transfer inherent in the line-formation process.

In this section we summarize the forward modeling of \ion{O}{6}
line profiles for an arbitrary set of coronal parameters ({\S}~3.1),
then we describe how these parameters are specified and varied to
produce various empirical model grids ({\S}~3.2).
Finally, we present a new method of computing the probability of
agreement between a given empirical model and the observations
({\S}~3.3), such that no regions of the possible solution space
are neglected.

\subsection{Forward Modeling}

The \ion{O}{6} line emission in coronal holes comes from two
sources of comparable magnitude:
(1) collisional electron impact excitation followed by
radiative decay, and (2) resonant scattering of photons that
originate on the bright solar disk.
The emergent specific intensity of an emission line from an
optically thin corona is given by
\begin{equation}
  I_{\nu} \, = \, \int_{-\infty}^{+\infty} dx \,
  \left( j_{\nu}^{\rm coll} + j_{\nu}^{\rm res} \right) \,\,\, ,
  \label{eq:Iline}
\end{equation}
where $x$ is the coordinate direction along the observer's
line of sight (LOS) and $j_{\nu}^{\rm coll}$ and
$j_{\nu}^{\rm res}$ are the collisionally excited and
resonantly scattered line emissivities, respectively.
We neglect the relatively weak UV continuum and the Thomson
electron-scattered components of the spectral lines in question.
At a given point in the three-dimensional coronal hole volume,
the line emissivities are specified by 
\begin{eqnarray}
  j_{\nu}^{\rm coll} &=& \frac{h\nu_{0}}{4\pi} \, q_{12}(T_{e}) \,
    n_{e} \, n_{1} \, \phi_{\nu} \label{eq:jcoll} \\
  j_{\nu}^{\rm res}  &=& \frac{h\nu_{0}}{4\pi} \, B_{12} \, n_{1}
    \int_{0}^{\infty} d\nu' \oint \frac{d\Omega'}{4\pi} \,
    R(\nu', {\bf\hat{n}}', \nu, {\bf\hat{n}}) \,
    \tilde{I}_{\nu'}({\bf\hat{n}}')
    \label{eq:jres}
\end{eqnarray}
(see, e.g., Mihalas 1978).
Here, $\nu_{0}$ is the rest-frame line-center frequency,
$q_{12}$ is the collision rate per particle for the transition
between atomic levels 1 and 2, $n_1$ is the number density
in the lower level of the atom or ion of interest (here, the
$2s$~$^{2} \! S_{1/2}$ state of O$^{5+}$),
and $B_{12}$ is the Einstein absorption rate of the transition.
The emission profile $\phi_{\nu}$ is assumed to be Gaussian.
The scattering redistribution function $R$ takes the incident
frequency $\nu'$ and photon direction vector ${\bf\hat{n}}'$ and
transforms it into the observed frequency $\nu$ along the LOS
direction ${\bf\hat{n}}$.

The profile $\phi_{\nu}$ and the redistribution function $R$
contain the main dependences on the properties of the ion
velocity distribution.
We allow for the possibility of an anisotropic O$^{5+}$ velocity
distribution by using a bi-Maxwellian function (e.g.,
Whang 1971), with the parallel and perpendicular axes oriented
arbitrarily with respect to the radial direction in the corona;
see {\S}~3.2.
The emissivity profiles along the LOS are modeled with the full
effects of the bi-Maxwellian velocity distribution and the
projected components of the bulk outflow speed along the
${\bf\hat{n}}'$ and ${\bf\hat{n}}$ directions.
For the polar coronal hole measurements being modeled here, we
define a Cartesian coordinate system for which the LOS direction
is denoted $x$ and the north-south polar axis of the Sun is $z$.
The other coordinate $y$ is set to zero.
General expressions for the emissivities are given in various
levels of detail by Withbroe et al.\  (1982),
Noci et al.\  (1987), Allen et al.\  (1998), Cranmer (1998),
Li et al.\  (1998), Noci \& Maccari (1999),
Kohl et al.\  (2006), and Akinari (2007).

The resonantly scattered components depend sensitively on
the intensity profiles incident from the solar disk
($\tilde{I}_{\nu'}$).
As in Cranmer et al.\  (1999), we used empirically derived
Gaussian profiles with total intensities measured on the
disk by UVCS at solar minimum (Raymond et al.\  1997).
The adopted \ion{O}{6} $\lambda$1032 (1031.93 {\AA}) disk
intensity is $1.94 \times 10^{13}$
photons s$^{-1}$ cm$^{-2}$ sr$^{-1}$, and the total intensities
of the \ion{O}{6} $\lambda$1037 (1037.62 {\AA}),
\ion{C}{2} $\lambda$1037.02, and \ion{C}{2} $\lambda$1036.34
disk lines are 0.500, 0.214, and 0.171 times the $\lambda$1032
intensity, respectively.
We used the profile widths as given by Noci et al.\  (1987);
see also the comparative tables of Gabriel et al.\  (2003)
and Raouafi \& Solanki (2004).

The collisional components depend on how the collision rate
$q_{12}$ varies with electron temperature $T_e$.
We kept the same tabulated values as were used by
Raymond et al.\  (1997) and Cranmer et al.\  (1999).
For completeness, we give a fit to $q_{12} (T_{e})$ for the
\ion{O}{6} $\lambda$1032 transition:
\begin{equation}
  \log_{10} (q_{12}) \, = \, -0.22117 t^{2} + 2.4565 t - 14.695
  \,\, ,
\end{equation}
where $t = \log_{10} T_{e}$, and $T_e$ and $q_{12}$ are given in
units of K and cm$^{3}$ s$^{-1}$ respectively.
This expression is valid to within about $\pm$ 2\% over the
range $5.3 \leq t \leq 6.3$.
The collision rate for the \ion{O}{6} $\lambda$1037 line is
half of that of the \ion{O}{6} $\lambda$1032 line.

The numerical code that synthesizes line profiles by numerically
integrating equations (\ref{eq:Iline})--(\ref{eq:jres}) is
essentially the same as the one used by Cranmer et al.\  (1999).
The integrations over $x$ and $\nu'$ have been simplified by
replacing the adaptive Romberg method by fixed grids, with
spacings that have been adjusted to minimize both numerical
discretization errors and run time.
The LOS integration was performed in steps of 0.1 $R_{\odot}$
from --15 to $+$15 $R_{\odot}$ along the $x$ axis.
The incident frequency grid corresponds to a wavelength grid
with a spacing of 0.03 {\AA} in $\lambda'$.
These step sizes were verified to give accurate results by
halving the step sizes and obtaining the same results to within
a desired precision.
We integrated over the solid angle of the solar disk
($d\Omega' = \sin\theta' \, d\theta' \, d\phi'$)
by Gauss-Legendre quadrature in $\theta'$ and equally spaced
trapezoidal quadrature in $\phi'$.
The solar disk was assumed to be uniformly bright.

\subsection{Parameter Selection for Line Synthesis}

For the \ion{O}{6} doublet, there are three primary observables
($I_{\rm tot}$ of $\lambda$1032, $V_{1/e}$, and ${\cal R}$)
that depend on the LOS distributions of four ``unknown''
quantities as well as a longer list of quantities that can be
considered to be known independently of the UVCS observations.
The four unknowns are the ion fraction (essentially $n_{1}/n_{e}$),
the O$^{5+}$ bulk outflow speed along the magnetic field
($u_{i \parallel}$), and the parallel and perpendicular O$^{5+}$
kinetic temperatures ($T_{i \parallel}$ and $T_{i \perp}$).
The known quantities include the electron density $n_e$,
the electron temperature $T_e$, the incident intensity from
the solar disk, and the overall magnetic geometry of the coronal
hole (i.e., how to compute ``parallel'' and ``perpendicular'' at
any point along the LOS).
Note that both emissivities
(eqs.\  [\ref{eq:jcoll}]--[\ref{eq:jres}]) depend linearly on the
ion fraction, so that the total intensity $I_{\rm tot}$ can
be used as a straightforward diagnostic of this quantity after
the other parameters have been determined.
The line widths and intensity ratios do not depend on the ion
fraction.
This leaves two observables ($V_{1/e}$ and ${\cal R}$) to
specify the values of three ion quantities
($u_{i \parallel}$, $T_{i \parallel}$, and $T_{i \perp}$).
Although this system is formally underdetermined, we can
nonetheless put some firm limits on the {\em ranges} of these
quantities and compute the most probable values.

Below, the three O$^{5+}$ velocity distribution parameters are
discussed in {\S}~3.2.1 and the other ``known'' parameters are
discussed in {\S}~3.2.2.

\subsubsection{Ionized Oxygen Parameters}

We treat the three unknown ion quantities as free parameters
that are varied independently of one another.
Other empirical modeling efforts (e.g., Cranmer et al.\  1999;
Antonucci et al.\  2000; Raouafi \& Solanki 2004, 2006) have tended
to use some form of iterative refinement; i.e., they started with
a specific set of initial conditions and assumptions, and they
varied some parameters---and kept others fixed---to find the most
probable values of $u_{i \parallel}$, $T_{i \parallel}$,
and $T_{i \perp}$.
The initial estimates tended to utilize the fact that the line
widths are most sensitive to $T_{i \perp}$, whereas the line
ratios depend mainly on the effect of Doppler dimming (and
Doppler pumping from the \ion{C}{2} solar disk lines) and
thus are sensitive mainly to the parallel velocity distribution
($u_{i \parallel}$ and $T_{i \parallel}$).
These iterative procedures contain the inherent possibility that
some regions of the parameter space could be neglected, and thus
possibly valid solutions could be ignored.
In this paper we {\em search the entire parameter space} by
constructing a three-dimensional ``data cube'' which contains
all possible combinations of the three parameters.

The three axes of the data cube were chosen to be
$u_{i \parallel}$, $T_{i \perp}$, and the anisotropy ratio
$T_{i \perp} / T_{i \parallel}$.
The modeled ranges of these quantities were made as wide as
possible in order to avoid missing possibly relevant regions of
parameter space.
The outflow speed $u_{i \parallel}$ was varied between 0 and
1000 km s$^{-1}$ using a linearly spaced grid.
The perpendicular kinetic temperature $T_{i \perp}$ was varied
logarithmically between $5 \times 10^{5}$ and $10^{9}$ K.
The anisotropy ratio was varied logarithmically between
0.1 and 100.
There were 50 values of each parameter along the three axes
of the data cube, and we synthesized 12 wavelengths---spaced
linearly between the line center and 2.7 {\AA} redward of line
center---for both \ion{O}{6} lines.
Thus, a data cube constructed for a specific height in the corona
($z$) consisted of $3 \times 10^6$ ($50^3 \times 24$) individual
LOS integrations.

For each point in a data cube, the scalar values of 
$u_{i \parallel}$ and $T_{i \perp}$ were assumed to be those
in the plane of the sky (i.e., $x=0$).
For other points along the LOS, the models used slightly larger
values that are consistent with an assumed radial increase in
both parameters.
Mass flux conservation---using the modeled $n_{e}(r)$ and flux
tube geometry---was used to specify the radial increase in
$u_{i \parallel}$ along the LOS.
Earlier empirical modeling results (specifically, eq.~[28]
of Cranmer et al.\  1999) were used to specify the radial
increase in $T_{i \perp}$.
The modeled anisotropy ratio $T_{i \perp} / T_{i \parallel}$
was assumed to remain constant along the LOS.
It is important to note that the modeled radial increases in
$u_{i \parallel}$ and $T_{i \perp}$ were always taken to be
{\em relative} to the plane-of-sky values that were varied
freely throughout each data cube.
Thus, there is no {\em a~priori} reason for the resulting
most-probable values of these parameters (determined via
comparisons with observations over a range of heights $z$)
to exhibit similar radial increases.\footnote{%
Because the degree of radial increase in $T_{i \perp}$ is
relatively uncertain, we constructed an additional set of empirical
models with no radial increase in $T_{i \perp}$ (i.e., where the
plane-of-sky values were kept constant over the LOS).
The resulting probability distributions ({\S}~4) were
virtually identical to those computed with the specified
radial increase along the LOS.}

We note that the kinetic temperature quantities $T_{i \perp}$
and $T_{i \parallel}$ may describe some combination of
``thermal'' microscopic motions and any unresolved bulk motions
due to waves or turbulence.
Thus, there is a further step of interpretation required after
the most likely values of these kinetic temperatures have been
derived from the empirical modeling process.
Making a definitive separation between the thermal and nonthermal
components of these temperatures is beyond the scope of this paper.
However, we can make some qualitative comments on the likely
ranges of magnitude of these two components based on recent
theoretical models of Alfv\'{e}n waves in coronal holes; see
{\S\S}~4.2 and 4.3.

Finally, the O$^{5+}$ ion fraction $n_{1} / n_{e}$ was kept at a
constant (and arbitrary) value in all of the models.
Comparisons between the observed and synthesized total
intensities were used to derive measurements of this ion
fraction in the polar coronal holes; see {\S}~4.5.

\subsubsection{Electron and Flux Tube Parameters}

The three main ``known'' parameters that are explored in the
models shown below (but kept constant over each data cube) are
the electron density $n_{e}(r)$, electron temperature $T_{e}(r)$,
and the macroscopic flux-tube geometry of the coronal hole.
Any other parameters that could be varied---e.g., the disk
intensities of the \ion{O}{6} and \ion{C}{2} lines---were kept
fixed at the values given above in {\S}~3.1.

Because one main purpose of this paper is to determine why
the results of Raouafi \& Solanki (2004, 2006) appear to differ
from earlier empirical modeling efforts, we constructed two
main sets of electron and flux tube parameters:
{\em model R,} which is designed to replicate many of the
conditions assumed by Raouafi \& Solanki (2004, 2006), and
{\em model C,} which is essentially the same as used by
Cranmer et al.\  (1999).
Below, we also discuss hybrid models with various combinations
of the conditions assumed in models R and C.

Model C uses an electron temperature derived by Ko et al.\  (1997)
from measurements of ion charge states in the fast solar wind
made by the SWICS instrument on {\em Ulysses}
(Gloeckler et al.\  1992).
We utilize the fitting formula
\begin{equation}
  T_{e} (r) \, = \, 10^{6} \, \mbox{K} \, \left[
  0.35 \left( \frac{r}{R_{\odot}} \right)^{1.1} +
  1.9  \left( \frac{r}{R_{\odot}} \right)^{-6.6} \right]^{-1}
  \,\, .
  \label{eq:Tko}
\end{equation}
For the electron density, model C uses the expression derived by
Cranmer et al.\  (1999) from direct inversion of UVCS/{\em{SOHO}}
white-light polarization brightness ($pB$) data over the poles
at solar minimum; i.e.,
\begin{equation}
  \frac{n_{e} (r)}{10^{5} \, \mbox{cm}^{-3}} \, = \,
  3890 \left( \frac{R_{\odot}}{r} \right)^{10.5} +
  8.69 \left( \frac{R_{\odot}}{r} \right)^{2.57}
  \,\, .
  \label{eq:neC99}
\end{equation}
The above electron density is a mean value for polar coronal
holes (intermediate between plumes and interplume regions)
between $r \approx 1.5$ and 4 $R_{\odot}$.
Model C also uses the three-parameter empirical function of
Kopp \& Holzer (1976) to specify the superradial expansion of a
polar coronal hole.
The transverse area $A(r) \propto r^{2} f(r)$ of the entire
coronal hole is specified by
\begin{equation}
  f(r) \, = \, 1 + (f_{\rm max} - 1) \left\{
  \frac{1 - \exp [( R_{\odot} - r) / \sigma_{1}]}
  {1 + \exp [( R_{1}     - r) / \sigma_{1}]} \right\} \,\, ,
  \label{eq:fofr}
\end{equation}
and Cranmer et al.\  (1999) determined
$f_{\rm max} = 6.5$, $R_{1} = 1.5 \, R_{\odot}$, and
$\sigma_{1} = 0.6 \, R_{\odot}$.
Also, the area of the hole is normalized by setting the
basal colatitude $\Theta_0$ to {28\arcdeg}.
The field lines inside the coronal hole volume are assumed to
self-similarly follow colatitudes that remain proportional to
the overall boundary of the coronal hole at any given radius
(see Cranmer et al.\  1999).

Model R uses a constant electron temperature of $10^6$ K.
This value is lower than the peak of the Ko et al.\  (1997)
model ($T_{e} \approx 1.5 \times 10^6$ K at
$r \approx 1.6 \, R_{\odot}$), and higher than the value
from this model at the coronal base ($T_{e} \approx
4 \times 10^{5}$ K at $r = R_{\odot}$).
The constant value of $10^6$ K seems to be in closer
agreement to both theoretical models that take account of
strong electron heat conduction in the corona (e.g.,
Lie-Svendsen \& Esser 2005; Cranmer et al.\  2007) and with
SUMER/{\em{SOHO}} observations made above the limb in
coronal holes (e.g., Wilhelm et al.\  1998;
Doschek et al.\  2001).\footnote{%
The discrepancies between electron temperatures derived from
spectroscopy and from frozen-in ion charge states are not yet
fully understood (see, e.g., Esser \& Edgar 2000, 2001;
Chen et al.\  2003; Laming \& Lepri 2007).}
For the electron density, model R uses equation (2) of
Doyle et al.\  (1999), i.e.,
\begin{equation}
  \frac{n_{e} (r)}{10^{5} \, \mbox{cm}^{-3}} \, = \,
  1000  \left( \frac{R_{\odot}}{r} \right)^{8} +
  0.025 \left( \frac{R_{\odot}}{r} \right)^{4} +
  2.9   \left( \frac{R_{\odot}}{r} \right)^{2}  \,\, .
  \label{eq:neD99}
\end{equation}
To specify the superradial geometry of flux tubes in the polar
coronal hole, model R uses the analytic magnetic field model of
Banaszkiewicz et al.\  (1998).

Note that Raouafi \& Solanki (2004, 2006) used equation (1)
of Doyle et al.\  (1999), which is a one-parameter hydrostatic 
fit to various measured electron densities.
Above a height of $r \approx 6 \, R_{\odot}$, though,
the radial decrease of $n_e$ in the hydrostatic expression
becomes substantially {\em shallower} than an inverse-square
radial decrease.
This is not generally expected to occur; i.e., in most
observations and models, the radial decrease in $n_e$ goes from
a rate much steeper than $1/r^2$ at low heights to $1/r^2$ at
large heights where the geometry is radial and the wind
speed is constant.
The shallow radial density decrease in a hydrostatic model
is probably unphysical and could lead to an overestimated
contribution from large distances along the LOS.

\begin{figure}
\epsscale{1.13}
\plotone{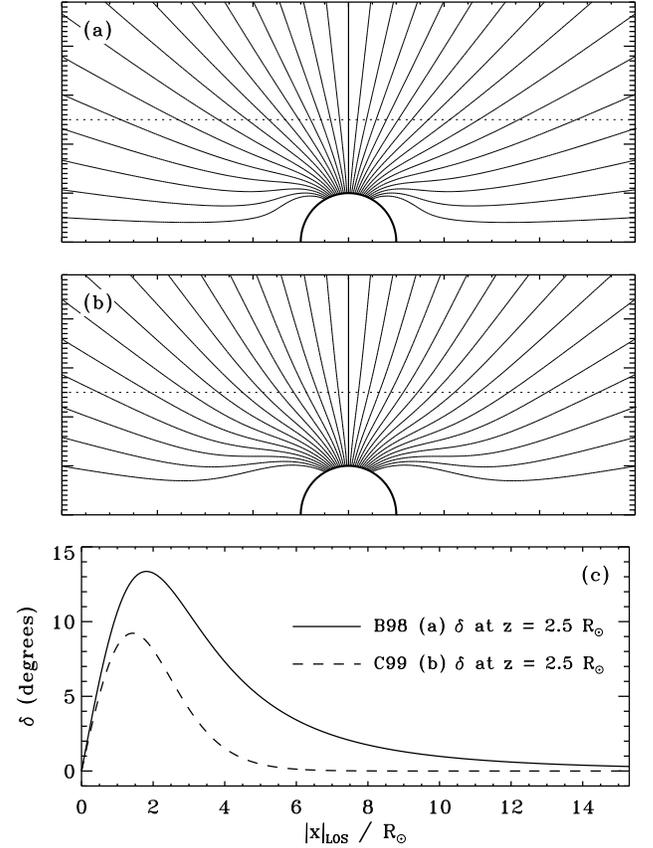}
\caption{Magnetic field lines in the plane of the sky for
{\em (a)} the B98 (Banaszkiewicz et al.\  1998) model, and
{\em (b)} the C99 (Cranmer et al.\  1999) model that used
the Kopp \& Holzer (1976) flux-tube area function.
The dotted horizontal line denotes the position of the LOS
along which various quantities are plotted in {\em (c)}.
In panel {\em (c)}, the superradial angle $\delta$ is given
as a function of $|x|$ (it is the same in the foreground and
background halves of the LOS) for the two models shown above
(see labels).}
\end{figure}

Figure 2 illustrates the differences between the magnetic
geometries used in models R and C.
Figures 2{\em{a}} and 2{\em{b}} show field lines that are
distributed evenly in polar angle $\theta$ between
{0\arcdeg} and {29\arcdeg} as measured on the solar surface
from the north pole.
The superradial angle $\delta$ characterizes the departure from
the radial direction, and it is shown in Figure 2{\em{c}} as a
function of LOS distance $x$ for a polar observing height
$z = 2.5 \, R_{\odot}$.
Formally, $\delta$ is defined as the angle between the radius
vector ${\bf r}$ and the magnetic field ${\bf B}$ (assuming the
field points outward), i.e.,
\begin{equation}
  \delta \, = \, \cos^{-1} \left(
  \frac{{\bf r} \cdot {\bf B}}{|{\bf r}| \, |{\bf B}|} \right)
  \,\, .
\end{equation}
For the polar observations described here, the LOS projection of
any quantity that follows the magnetic field (e.g., the outflow
velocity) is given by multiplying its magnitude by
$\sin (\theta + \delta)$.
The Banaszkiewicz et al.\  (1998) model exhibits a larger
degree of departure from radial geometry than does the
Cranmer et al.\  (1999) model.
However, at the large heights for which the UVCS \ion{O}{6}
anisotropy results are of interest here, the relative differences
between the two models---and also the differences between either
model and a radial geometry ($\delta = 0$)---are small.

Figure 3 shows the range of electron densities measured by
several instruments in polar coronal holes.
The strong radial decrease in $n_{e}(r)$ has been removed by
dividing all measurements by equation (\ref{eq:neD99}).
The use of this normalization more clearly illustrates the
relative differences between the different sets of values, which
Raouafi \& Solanki (2004, 2006) claimed to be important in the
derivation of O$^{5+}$ temperature anisotropy.
The differences between plumes and interplume regions is certainly
responsible for some of the wide range of variation, but some of
it may also be due to absolute calibration uncertainties between
instruments.
Note, though, that the curve representing the hydrostatic
equation (1) of Doyle et al.\  (1999) appears to be clearly
divergent from the other empirical curves above 
$r \approx 8 \, R_{\odot}$, with a slope that is flatter than the
other measurements even several solar radii below that.

\begin{figure}
\epsscale{1.13}
\plotone{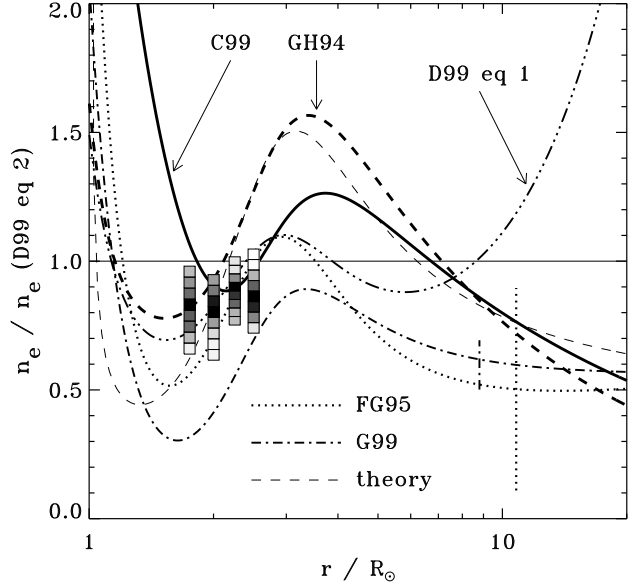}
\caption{Comparison of measured electron densities in polar
coronal holes.  Values of $n_e$ from
Guhathakurta \& Holzer (1994) ({\em thick dashed line}),
Fisher \& Guhathakurta (1995) ({\em dotted line \& vertical bar}),
Guhathakurta et al.\  (1999) ({\em dot-dashed line \& vertical
bar}), and Cranmer et al.\  (1999) ({\em thick solid line})
were divided by equation (\ref{eq:neD99}), i.e.,
equation (2) of Doyle et al.\  (1999).
The polar theoretical model of Cranmer et al.\  (2007)
is also shown ({\em thin dashed line}), as is the approximate
hydrostatic fit from eq.~(1) of Doyle et al.\  (1999)
({\em triple-dot-dashed line}).
Gray-scale histogram boxes show the range of $n_e$ values from
the plume statistics study of Cranmer et al.\  (1999), with
darker shades denoting the most likely values at each height.}
\end{figure}

The gray-scale histogram boxes in Figure 3 illustrate the
variations due to differing concentrations of polar plumes along
a single polar LOS over three months (1996 November 1 to
1997 February 1) using a consistent data set and large enough
count rates to make Poisson uncertainties negligible
(see Table 3 of Cranmer et al.\  1999).
The curves from Fisher \& Guhathakurta (1995) and
Guhathakurta et al.\  (1999) show averages of the various
plume and interplume values given in those papers, with
vertical lines illustrating the relative contrast between the
densest plume-filled lines of sight and the regions with the
fewest numbers of plumes.
(These vertical lines are shown at $r \approx 10 \, R_{\odot}$
for clarity, but they are representative of the values at the
lower heights corresponding to the observed white-light data.)
Polar electron density values reported recently by
Qu\'{e}merais et al. (2007) are not shown, but they are similar
in radial shape to the Fisher \& Guhathakurta (1995) mean curve
(but with values about 10\%--20\% higher).
Overall, the variations between data sets that appear to exceed
the plume-interplume contrast may be due to different instrument
calibrations.

When modeling the \ion{O}{6} observations summarized in {\S}~2,
it is probably incorrect to use the lowest ``pure interplume''
electron densities.
At $z \approx 2.5$--3 $R_{\odot}$ in polar coronal holes, the 
UVCS observations were typically integrated over {15\arcmin} to
{30\arcmin} in the tangential direction, whereas polar plumes
at these heights have transverse sizes of only about {1\arcmin}
to {2\arcmin}.
Thus, the most appropriate electron densities to use are those
that {\em average} over plumes and interplume regions.
The lower limits from Fisher \& Guhathakurta (1995)
and Guhathakurta et al.\  (1999), as well as the fitting function
given by Esser et al.\  (1999), seem to be inappropriate to apply
to the empirical modeling of these UVCS \ion{O}{6} data.
At lower heights, where the plume and interplume regions have
been resolved by UVCS (e.g., Kohl et al.\  1997;
Giordano et al.\  2000), the use of the full range of plume and
interplume values of $n_e$ would be warranted.

\subsection{Comparison with Observations}

Once a model data cube (which varies $u_{i \parallel}$,
$T_{i \perp}$, and $T_{i \perp} / T_{i \parallel}$ along its
axes) has been produced for a given observing height $z$ and
a given set of $n_e$, $T_e$, and flux tube parameters, the
next step is to compute the probability of agreement between
a given observation and each of the simulated observations
in the cube.
We compute this probability $P$ as the product of two quantities
that are assumed to be independent of one another:
(1) the probability $P_{\rm R}$ that the observed line ratio
agrees with the simulated ratio, and (2) the probability
$P_{\rm S}$ that the observed profile shape of the
\ion{O}{6} $\lambda$1032 line agrees with the simulated shape.
Because the brighter \ion{O}{6} $\lambda$1032 line tends to
dominate the measured ``weighted'' line width $V_{1/e}$, we
use only the simulated \ion{O}{6} $\lambda$1032 line shape in
the latter comparison.

The line ratio probability $P_{\rm R}$ is relatively
straightforward to compute.
The modeled total intensities of the two components of the
doublet are determined by summing up the specific intensities
over the 12 wavelength bins.
Their ratio thus gives ${\cal R}_{\rm model}$.
The relative distance between ${\cal R}_{\rm model}$ and the
observed ratio ${\cal R}_{\rm obs}$, in units of the
observational standard deviation ($\delta {\cal R}_{\rm obs}$),
is the quantity that determines the probability of agreement.
Assuming the uncertainties are normally distributed, the
probability is
\begin{equation}
  P_{\rm R} \, = \, 1 - \mbox{erf} \left(
  \frac{|{\cal R}_{\rm obs} - {\cal R}_{\rm model}|}
  {\delta {\cal R}_{\rm obs} \, \sqrt{2}} \right)
  \label{eq:PR}
\end{equation}
(see, e.g., Bevington \& Robinson 2003).
A larger argument in the error function (``erf'' above) denotes
a larger discrepancy between the modeled and observed ratios,
and thus a lower probability of agreement.

The line shape probability $P_{\rm S}$ is not as easy to
compute as $P_{\rm R}$.
An initial attempt was made to fit the simulated profiles with
Gaussian functions, and then to compare the resulting $V_{1/e}$
widths with the observed values using a similar expression as
equation (\ref{eq:PR}).
However, there were many instances where the modeled lines were
far from Gaussian in shape, but the best-fitting Gaussian (which
was a poor fit in an absolute sense) happened to agree with the
observed $V_{1/e}$.
This resulted in spuriously high probabilities for wide regions
of parameter space that should have been excluded.
Thus, we found that the tabulated specific intensities (i.e.,
the full {\em line shapes}) need to be compared on a
wavelength-by-wavelength basis.
This raises the issue of what to use for the ``observed'' line
shape.
As described in {\S}~2, the UVCS/{\em{SOHO}} data points contain
a wide range of instrumental effects that were taken into account
in the line fitting process.
In order to compare similar quantities, either these effects must
be inserted into the model profiles, or we must reconstruct
``observed profile'' information from the extracted $V_{1/e}$
measurements and the $\delta V_{1/e}$ uncertainties.
We chose the latter option.

To determine the probability of agreement between the set of
modeled specific intensities ($I_{\lambda, {\rm model}}$) and
the reconstructed observed intensities ($I_{\lambda, {\rm obs}}$),
we computed a $\chi^2$ quantity,
\begin{equation}
  \chi^{2} \, = \, \sum_{\lambda} \left(
  \frac{I_{\lambda, {\rm obs}} - I_{\lambda, {\rm model}}}
  {\delta I_{\lambda, {\rm obs}}}
  \right)^{2}
\end{equation}
where $I_{\lambda, {\rm model}}$ came from the data cube, and
$I_{\lambda, {\rm obs}}$ was constrained to be a Gaussian
function with the observed $V_{1/e}$ width and a total intensity
equal to that of the modeled profile.
(The observed total intensity was not used because the comparison
being done here is only between the relative shapes.)
The $\delta I_{\lambda, {\rm obs}}$ uncertainty was computed as
a function of wavelength by comparing the idealized
$I_{\lambda, {\rm obs}}$ profile with two others computed with
line widths $V_{1/e} - \delta V_{1/e}$ and
$V_{1/e} + \delta V_{1/e}$ (with all three profiles normalized
to the same modeled total intensity).
These three profiles exhibited a range of specific intensities
at each wavelength, and the standard deviation quantity
$\delta I_{\lambda, {\rm obs}}$ was defined as half of that
full range.
Then, the $\chi^{2}$ quantity above constrains the probability
that the observed and modeled profiles are in agreement
(i.e., the probability that the observed and modeled specific
intensity values are drawn from the same distribution).
Assuming normally distributed uncertainties, this probability
is given by
\begin{equation}
  P_{\rm S} \, \equiv \, Q (\chi^{2} | \nu)
  = \frac{1}{\Gamma (\nu / 2)}
  \int_{\chi^{2} / 2}^{\infty} e^{-t} t^{(\nu / 2)-1} dt
\end{equation}
(Press et al.\  1992),
where $\nu = N_{\lambda}-1$ is the effective number of degrees
of freedom (for $N_{\lambda} = 12$ wavelength points) and
$\Gamma(x)$ is the complete Gamma function.
When $\chi^{2} \ll \nu$ the above probability
approaches unity (i.e., the modeled profile is a good
match to the observed profile), and when $\chi^{2} \gg \nu$
the above probability is negligibly small.

We thus obtained the total probability $P = P_{\rm R} P_{\rm S}$
as a function of the three main O$^{5+}$ variables of each
data cube, for each observation at the height $z$ consistent
with that data cube.
The question of what is considered to be a large or small
probability is open to some interpretation.
Below, we often use a standard ``one sigma'' probability
$P_{1\sigma} = 1 - \mbox{erf} (1/\sqrt{2}) = 0.317$ as a
fiducial value above which the solutions are considered to be
good matches with the data.

\section{Empirical Model Results}

In this section we present results for the most probable values of
the O$^{5+}$ ion properties between 1.5 and 3.5 $R_{\odot}$.
In {\S}~4.1, we show how the essential information inside the
three-dimensional probability cubes can be extracted and analyzed
in a manageable way.
In {\S}~4.2, the optimal values for O$^{5+}$ outflow speed and
perpendicular temperature are presented for models C and R.
The resulting values of $u_{i \parallel}$ and $T_{i \perp}$
are consistent with earlier determinations of preferential
ion heating and acceleration with respect to protons.
In {\S}~4.3, we discuss the determination of the anisotropy ratio
$T_{i \perp}/T_{i \parallel}$ for models C and R, which is less
certain than the other two quantities.
We then focus in detail on a single representative height in
{\S}~4.4 in order to determine how models C and R can give rise
to qualitatively different conclusions about the ion temperature
anisotropy.
Finally, in {\S}~4.5 we extract information from both
models C and R about the O$^{5+}$ ion concentration in the
extended corona---i.e., we compute the ratio
$n_{{\rm O}^{5+}} / n_{\rm H}$ from the comparison of observed
and modeled \ion{O}{6} $\lambda$1032 total intensities.

\subsection{Deriving Ion Properties from the Data Cubes}

We constructed two sets of radially dependent data cubes:
one for the model C assumptions for $n_e$, $T_e$, and flux-tube
geometry, and one for model R.
Each set consisted of 13 data cubes with observation heights
$z = 1.5$, 1.6, 1.7, 1.8, 1.98, 2.11, 2.3,
2.42, 2.563, 2.7, 3.0, 3.09, and 3.565 $R_{\odot}$.
These values were chosen to align with the observed data
points shown in Figure 1.
Any discrepancies between the observed and modeled heights
never exceeded $\pm 0.065 \, R_{\odot}$, and for the whole
data set the average absolute value of the discrepancy was
only 0.012 $R_{\odot}$.
We then constructed 49 individual ``probability cubes'' for
each of the data points for which both $V_{1/e}$ and
${\cal R}$ exist.

\begin{figure}
\epsscale{1.13}
\plotone{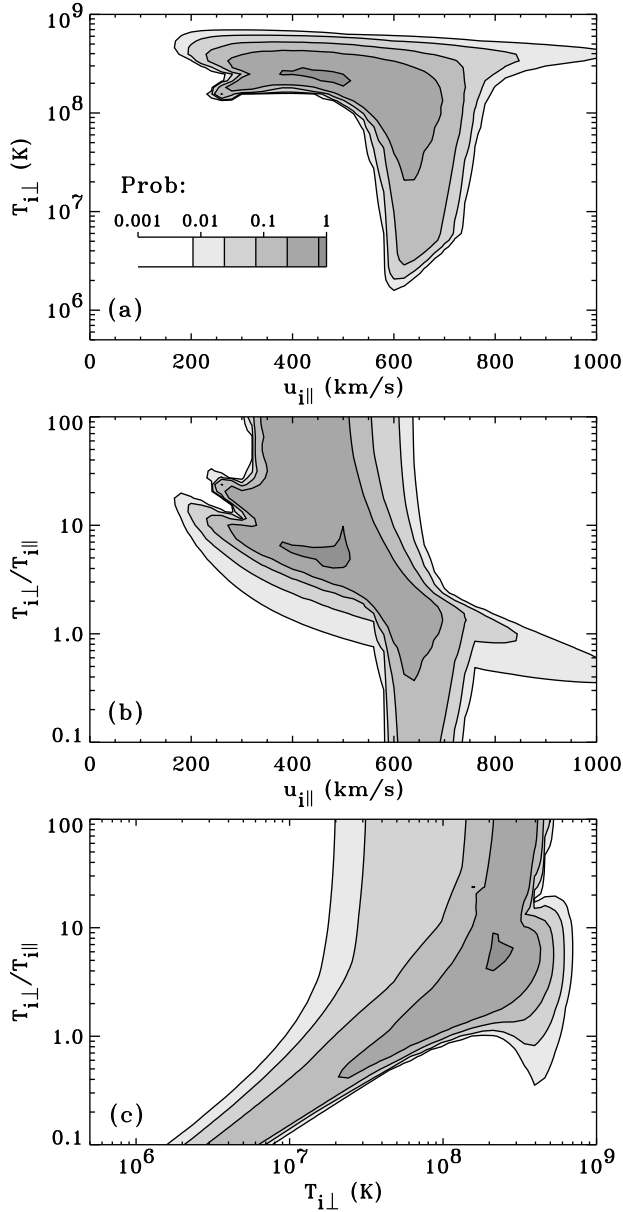}
\caption{Contour plots of the maximum probabilities of agreement
between the model R data cube and the UVCS observation from
1997 January 5 at $r = 3.07 \, R_{\odot}$.
The three panels show probabilities as a function of each
unique pair of the three data-cube axis quantities.
Contour levels are plotted at 90\% of the probability
values 1, 0.3, 0.1, 0.03, and 0.01 (see gray-scale code in
panel {\em a}).}
\end{figure}

Even for just a single comparison between an observation and a
data cube at one height, it is a challenge to display the
full three-dimensional nature of the probability ``cloud''
$P(u_{i \parallel}, T_{i \perp}, T_{i \perp}/T_{i \parallel})$.
We limit ourselves to showing lower-dimensional projections
that keep only the highest probability values taken over the
axes that are not being shown.
As an example, in Figure 4 we display two-dimensional contours
of $P$ as a function of all three unique pairings of the three
axis-quantities of the data cube.
The specific comparison is between the measurement shown in
Table 1 from 1997 January 5 ($3.07 \, R_{\odot}$,
$V_{1/e} = 690$ km s$^{-1}$) and the model R data cube
constructed at $z = 3.09 \, R_{\odot}$.
In all three contour plots, the probability shown at each
location is a maximum taken over the third quantity that is
orthogonal to the projection plane.
Thus, for regions with a low probability in these diagrams,
we can be assured that there are {\em no} values of the unplotted
coordinate that can give synthetic line profiles in agreement
with the observations.
 
Figure 4{\em{a}} shows an approximate anticorrelation between
the ion outflow speed and the perpendicular kinetic temperature
in the subset of generally ``successful'' models.
This arises mainly because the lines can be broadened both
by microscopic LOS motions (roughly proportional to
$T_{i \perp}$) and by the projection of the superradially
flowing bulk outflow speed along the LOS (which goes as
$u_{i \parallel}$).
When one of these quantities goes up, the other must go down
in order to match a given observed line profile.
Figure 4{\em{b}} shows that the region of parameter space
with the larger contribution by $T_{i \perp}$ (upper left)
also requires a large anisotropy ratio, but the region with
the larger contribution by bulk LOS motions (lower right)
may be able to match the observations with an isotropic
velocity distribution (i.e.,
$T_{i \perp}/T_{i \parallel} \approx 1$).

The large amount of information in contour plots like Figure 4
can be collapsed down to a smaller list of parameters.
We created three one-dimensional probability curves as a
function of each of the three main axis quantities, with the
maximum values extracted from the full plane subtended by the
remaining two neglected quantities.
Thus, we define the reduced probability functions
$P_{u} (u_{i \parallel})$, $P_{t} (T_{i \perp})$, and
$P_{a} (T_{i \perp}/T_{i \parallel})$ (the subscript ``$a$''
denotes the anisotropy ratio).
These functions are generally peaked at some high value close
to 1 and exhibit lower values far from the optimal solutions.
Figure 5 shows these reduced probabilities for the same example
case shown in Figure 4.
The reduced probabilities for $u_{i \parallel}$ and $T_{i \perp}$
are peaked relatively sharply around their most probable values.
Note that we plot the perpendicular kinetic temperature in units
of a most-probable speed
$w_{i \perp} = (2 k_{\rm B} T_{i \perp} / m_{i})^{1/2}$
in order to facilitate comparison with earlier papers.
The reduced probability for the anisotropy ratio, shown
in Figure 5{\em{b}}, is less centrally peaked and thus the best
solution for this value is less certain.
The peak value corresponds to a most-probable anisotropy
ratio of $T_{i \perp}/T_{i \parallel} \approx 6$, but note that
the probability of isotropy remains reasonably high at
$\sim$40\%.

\begin{figure}
\epsscale{1.13}
\plotone{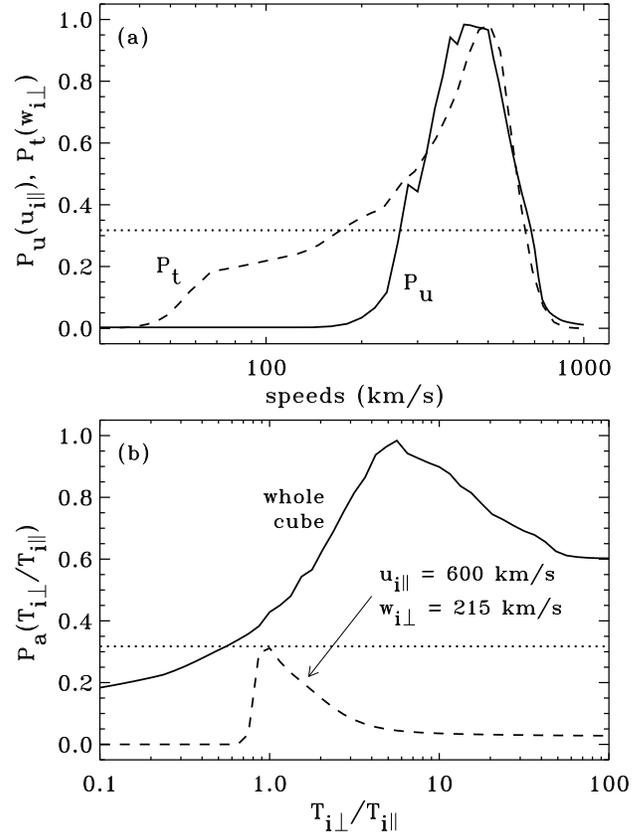}
\caption{Reduced probabilities for one specific comparison
between a UVCS observation at $r = 3.07 \, R_{\odot}$ and the
model R data cube (see also Figs.\  4, 8, 10, and 11).
({\em a}) $P_u$ versus outflow speed $u_{i \parallel}$
({\em solid line}) and $P_t$ versus perpendicular most-probable
speed $w_{i \perp}$ ({\em dashed line}).
({\em b}) $P_a$ versus anisotropy ratio
$T_{i \perp}/T_{i \parallel}$ for a search of the entire data
cube ({\em solid line}) and for a ``slice'' through the data
cube with fixed values of $u_{i \parallel}$ and $w_{i \perp}$
given above ({\em dashed line}).
Also shown is the threshold level $P_{1\sigma}$
({\em dotted lines}) defined in the text.}
\end{figure}

It is interesting to contrast the exhaustive data-cube-search
technique used here with the more straightforward approaches
taken in earlier papers.
For example, Raouafi \& Solanki (2004, 2006) simulated the
properties of the \ion{O}{6} $\lambda\lambda$1032, 1037 lines
after first fixing the radial variation of the outflow speed
and ion temperature.
Figure 5{\em{b}} shows an illustrative ``cut'' through the
data cube at {\em fixed} values of $u_{i \parallel} = 600$
km~s$^{-1}$ and $w_{i \perp} = 215$ km~s$^{-1}$ at
$r = 3.09 \, R_{\odot}$ (similar to the values used by
Raouafi \& Solanki 2006).
In this case, the most probable value of the anisotropy ratio is
surprisingly close to 1, as was also assumed by Raouafi \& Solanki.
The apparent consistency with the observations (i.e., a value of
$P_a$ of about 30\%) may be misleading if the rest of the data
cube is not searched.
Thus, we can assert that any results concerning the anisotropy
ratio that were obtained by {\em not} searching the full range
of possibilities for the ion parameters are potentially
inaccurate.

The ultimate goal of the empirical modeling process is to
characterize the peak values and widths of the reduced probability
curves, in order to obtain the optimal measured values (with
uncertainty limits) for the relevant O$^{5+}$ plasma properties.
The most satisfactory outcome, of course, would be very narrow
peaks that occur far from the edges of the parameter space, but
this is not always the case.
After some experimentation, we chose to use weighted means
to obtain the peak values, i.e.,
\begin{equation}
  \langle x \rangle \, = \,
  \frac{\int dx \, x \, P_{x}(x)}{\int dx \, P_{x}(x)}
  \label{eq:Pmom}
\end{equation}
where $x$ denotes any of the three axis quantities $u_{i \parallel}$,
$T_{i \perp}$, or $\log (T_{i \perp}/T_{i \parallel})$.
We used the logarithm of the anisotropy ratio in
equation (\ref{eq:Pmom}) because tests showed that if the ratio
itself (which spans three orders of magnitude) was used, the
above mean would be weighted strongly toward the largest values
even when the maximum of the probability distribution is at much
lower values.

We experimented with using the variance, or second moment, of the
reduced probability distributions to characterize the widths of
``error bars'' for each measurement.
However, because the probability curves are generally not symmetric
around the peak values, the second moment often did not accurately
give a range of values with reasonable probabilities.
Instead, we performed a straightforward search for the range of
probabilities that are higher than the threshold value
$P_{1\sigma} \approx 0.317$ discussed above.
The lower and upper limits of that range were taken to be the
ends of the uncertainty bounds for each measurement.

\subsection{Preferential Ion Heating and Acceleration}

\begin{figure}
\epsscale{1.13}
\plotone{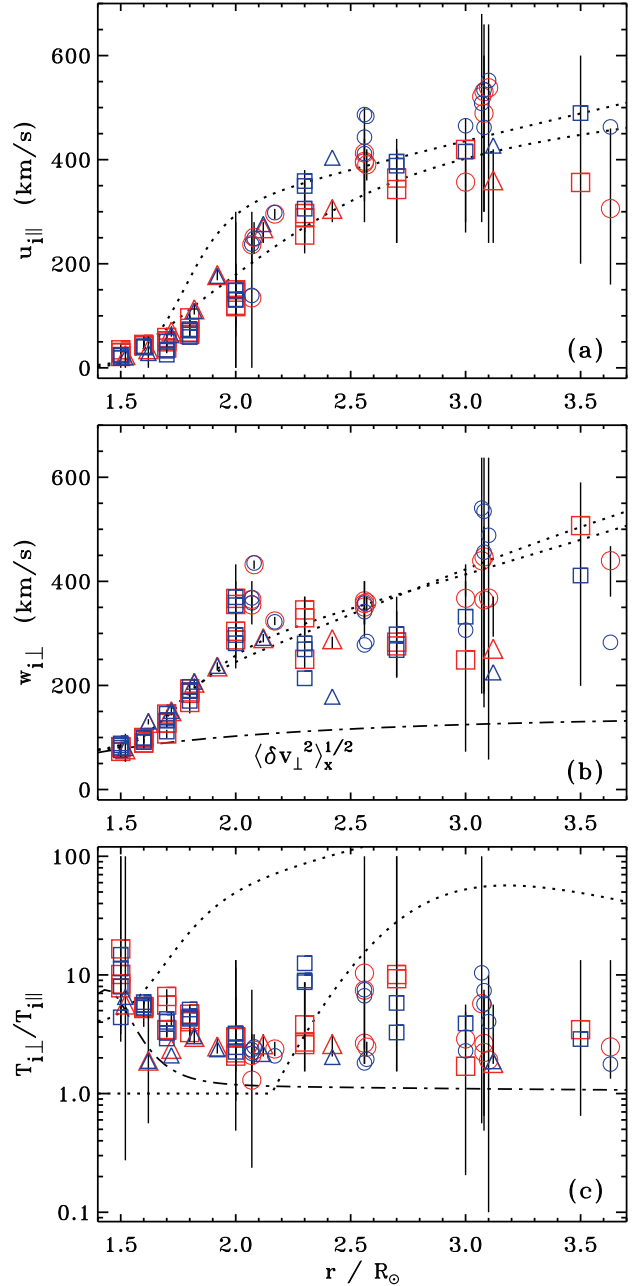}
\caption{Derived outflow speeds ({\em a}), perpendicular
most-probable speeds ({\em b}), and kinetic anisotropy ratios
({\em c}) for model R ({\em red points}) and model C
({\em blue points}).
Symbols show the weighted means of the reduced probability
distributions, with styles the same as in Figure 1.
Vertical bars show the full range of parameter space with
reduced probabilities greater than $P_{1\sigma}$ (for model R).
Also shown are empirical models B1 and B2 from
Cranmer et al.\  (1999) ({\em dotted lines}) and
Alfv\'{e}n wave quantities
$\langle \delta v_{\perp}^{2} \rangle_{x}^{1/2}$ in panel ({\em b})
and $A_{\rm eff}$ in panel ({\em c}), derived from the model
of Cranmer et al.\  (2007) ({\em dot-dashed lines}).}
\end{figure}

Figure 6 shows the weighted mean and error-bar quantities for
the O$^{5+}$ plasma properties, defined as in the previous
subsection, as a function of heliocentric height.
The results from model C and model R are plotted in two
different colors, with only the error bars of model R shown
for clarity.
Here we focus on the ion outflow speed (Fig.\  6{\em{a}})
and the perpendicular ion temperature (Fig.\  6{\em{b}}).
On average, the derived values of $\langle u_{i \parallel} \rangle$
and $\langle w_{i \perp} \rangle$ were consistent between the
two sets of models.
To quantify the impact of varying the electron density,
electron temperature, and flux-tube geometry, we computed ratios
of the model C values to the model R values for each data point.
For the 49 data points taken together, the mean value of the
ratio of the outflow speeds was 1.002, with a standard deviation
of 19\%, and the mean value of the ratio of perpendicular
most-probable speeds was 0.991, with a standard deviation of
14\%.
This shows that the determination of these parameters is
relatively insensitive to the choices of electron density,
electron temperature, and flux-tube geometry.

The radial dependence of the derived
$\langle u_{i \parallel} \rangle$ values in Figure 6{\em{a}}
is similar to that of the O$^{5+}$ empirical models B1 and B2
given by Cranmer et al.\  (1999).
Note the emergence of a natural trend of radial
acceleration in $\langle u_{i \parallel} \rangle$, with the
possible exception of the data points at
$r \gtrsim 3.5 \, R_{\odot}$.
This is especially serendipitous given that each data point
was analyzed independently of all others.

The derived O$^{5+}$ outflow speeds support earlier claims of
{\em preferential ion acceleration} in coronal holes.
At $r = 2.5 \, R_{\odot}$, the range of ion outflow speeds
that gives rise to high probabilities of agreement with the
data points is approximately 280--500 km s$^{-1}$ 
At this height, these values are substantially larger than
bulk (proton-electron) solar wind outflow speeds derived via
mass flux conservation.
Figure 41 of Kohl et al.\  (2006) showed a selection of 12
bulk outflow speed models derived using all possible
combinations of four $n_e$ models and three coronal-hole
geometries.
At $r = 2.5 \, R_{\odot}$, these 12 models gave a range of
bulk outflow speeds of 115--300 km s$^{-1}$.
Despite the small degree of overlap between the two ranges,
the mean value of the O$^{5+}$ range (390 km s$^{-1}$) 
exceeds the mean value of the mass flux conservation range
(208 km s$^{-1}$) by almost a factor of two.
Also, proton outflow speeds derived from \ion{H}{1} Ly$\alpha$
Doppler dimming (from a selection of papers all dealing with
polar coronal holes at the 1996--1997 solar minimum)
were shown in Figure 41 of Kohl et al.\  (2006).
At 2.5 $R_{\odot}$, the range of these values is
160--260 km s$^{-1}$---with a mean of 210 km s$^{-1}$---which
is still significantly lower than the range of O$^{5+}$ ion
outflow speeds discussed above.

In Figure 6{\em{b}}, the trend of radial increase in
$\langle w_{i \perp} \rangle$ is also roughly similar to that found
by Cranmer et al.\  (1999), especially below about 2.3 $R_{\odot}$.
At larger heights, though, there appears to be less evidence for
a systematic increase than existed in the model B1 and B2 curves.
This could have resulted either from the inclusion of the new
data points or from the more exhaustive treatment of uncertainties
in the new parameter determination method described above.
However, if one takes all of the derived $\langle w_{i \perp} \rangle$
values for $r \geq 2 \, R_{\odot}$ and fits to a straight line,
the best-fitting slope is still increasing with height at a rate
of 50 km s$^{-1}$ per $R_{\odot}$.
This is about a third of the $\sim$150 km s$^{-1}$ per $R_{\odot}$
slope in the B1 and B2 models.

The most-probable speeds $\langle w_{i \perp} \rangle$ shown in
Figure 6{\em{b}}, although slightly smaller than those given by
the Cranmer et al.\  (1999) model B1 and B2 curves at some heights,
still show definite evidence for {\em preferential ion heating.}
The mean value of the $\langle w_{i \perp} \rangle$ values at heights
$r \geq 2.5 \, R_{\odot}$ in Figure 6{\em{b}} is 363 km s$^{-1}$,
with a standard deviation of 73 km s$^{-1}$.
Between 2.5 and 3 $R_{\odot}$, the perpendicular proton
most-probable speeds derived from \ion{H}{1} Ly$\alpha$ were
about 210--240 km s$^{-1}$, with a mean value of about
225 km s$^{-1}$ (see models A1 and A2 of Cranmer et al.\  1999).
The fact that the O$^{5+}$ mean value exceeds the proton mean
value by almost two standard deviations implies that the
O$^{5+}$ kinetic temperature at this height is very likely to be
more than ``mass proportional'' (i.e., implying an oxygen kinetic
temperature of 130 MK, or more than 40 times the proton kinetic
temperature of $\sim$3 MK.

It is important to note that the derived kinetic temperatures are
likely to be a combination of thermal and nonthermal motions.
The ion-to-proton kinetic temperature ratio of $\sim$40, derived
above, is likely to be a {\em lower limit} to the true ratio of
thermal, or microscopic temperatures.
If unresolved wave motions are deconvolved from the empirical
values of $\langle w_{i \perp} \rangle$, the proton
most-probable speed will be reduced by a relatively larger
amount than the O$^{5+}$ speed.
As an example, the theoretical polar coronal hole model of
Cranmer et al.\  (2007) has a LOS-projected Alfv\'{e}n wave
amplitude at $r = 2.5 \, R_{\odot}$ of 116 km s$^{-1}$ (this is
also plotted in Fig.\  6{\em{b}}).
Converting these motions into temperature-like units and
subtracting from both values given above, one obtains an
O$^{5+}$ perpendicular temperature of 115 MK and a proton
perpendicular temperature of 2.2 MK.
The ratio of ion to proton temperatures has thus increased
from about 40 to 50.
In any case, it is clear that the dominant contributor to the ion
kinetic temperature is the true ``thermal'' temperature, with only
a relatively minor impact from broadening due to macroscopic motions.

\subsection{The Ion Anisotropy Ratio}

Figure 6{\em{c}} illustrates the largest discrepancy between
the empirical models of Cranmer et al.\  (1999) and the present
models (both C and R).
Above a height of $\sim$2.5 $R_{\odot}$, models B1 and B2
demanded a strong anisotropy ratio 
$T_{i \perp}/T_{i \parallel} > 10$, but the optimal ratios
derived in this paper seem to cluster between 2 and 10 with no
discernible radial dependence.
It is important to note, though, that there is considerable
overlap of the {\em uncertainties} between the old and new ranges
of $\langle T_{i \perp} / T_{i \parallel} \rangle$.
Several of the error bars shown in Figure 6{\em{c}}
extend up into the range of ratios from models B1 and B2.
Also, the dotted curves that illustrate models B1 and B2
correspond to the ``optimal'' values of the anisotropy ratio
from Kohl et al.\  (1998) and Cranmer et al.\  (1999); the
uncertainties in those models are not shown.

Because the derived values of the kinetic temperature ratio
$\langle T_{i \perp} / T_{i \parallel} \rangle$ in Figure 6{\em{c}}
exceed unity by only a relatively small amount, it is worthwhile to
examine whether the numerator ($T_{i \perp}$) may have been
enhanced by unresolved wave motions perpendicular to the
magnetic field.
In other words, for a realistic model of perpendicular wave
amplitudes in polar coronal holes, we investigate whether a
truly {\em isotropic} microscopic velocity distribution could
have given an effective anisotropy ratio that exceeds 1.
We compute such an effective anisotropy ratio as
\begin{equation}
  A_{\rm eff} \, = \, \frac{1}{1 - (\langle
  \delta v_{\perp}^{2} \rangle_{x} / w_{i \perp}^{2})}
  \,\, ,
\end{equation}
where $\langle \delta v_{\perp}^{2} \rangle_{x}$ is the
square of the frequency-integrated Alfv\'{e}n wave amplitude
divided by two to sample only the motions along one of the two
perpendicular directions (i.e., only along the LOS or $x$ axis).
As above, we used the Alfv\'{e}n wave properties from the
turbulence-driven polar coronal hole model of
Cranmer et al.\  (2007).
The model wave amplitude is plotted in Figure 6{\em{b}} and
the quantity $A_{\rm eff}$ is plotted in Figure 6{\em{c}}.
The condition $A_{\rm eff} \approx 1$ corresponds to the situation
where the amplitudes are too small to contribute to the
anisotropy ratio (as defined in the empirical models).
Below $r \approx 1.7 \, R_{\odot}$, the curve in Figure 6{\em{c}}
shows that $A_{\rm eff}$ does indeed exceed 1 by about the amount
computed from the UVCS data.
At these low heights, the derived value of
$\langle w_{i \perp} \rangle$ is of the same order of magnitude as
the wave amplitude, so the latter can ``contaminate'' the
determination of the true perpendicular most-probable speed.
At heights larger than about 2 $R_{\odot}$, though, the
wave amplitudes are small in comparison to the derived
$\langle w_{i \perp} \rangle$ values, and thus
$A_{\rm eff} \approx 1$.
We thus conclude that above 2 $R_{\odot}$, any derived
anisotropy ratio $\langle T_{i \perp} / T_{i \parallel} \rangle$
is likely to be truly representative of the microscopic velocity
distribution and not affected by wave motions.

Despite the comparatively low values of the anisotropy ratio
shown in Figure 6{\em{c}}
($\langle T_{i \perp}/T_{i \parallel} \rangle \approx 2$--10),
we should emphasize that these values are often significantly
different from unity.
It is important to note that {\em all} of the data points have
their largest reduced probability---measured either using the
weighted mean defined above or by simply locating the maximum
value---for anisotropy ratios larger than unity.

\begin{figure}
\epsscale{1.13}
\plotone{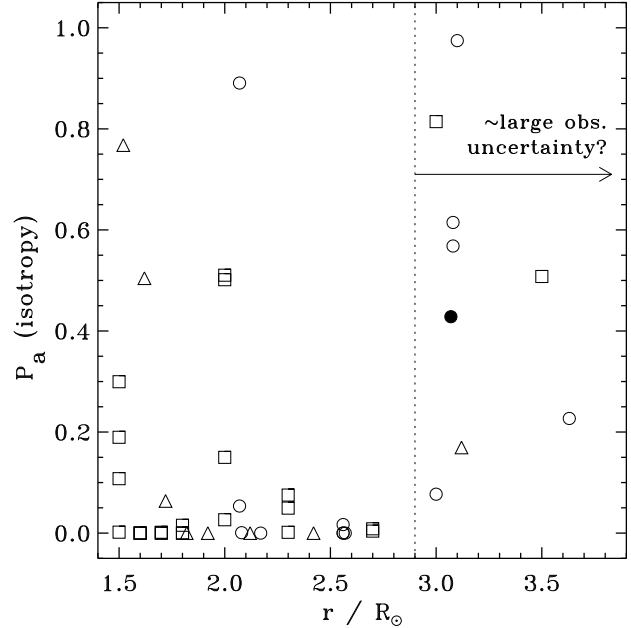}
\caption{Probability of ion isotropy $P_{a}(1)$ plotted
versus heliocentric distance for model R, using the same data
symbols as Figures 1 and 6.
The filled circle shows the probability of isotropy for the
specific observation ($3.07 \, R_{\odot}$,
$V_{1/e} = 690$ km s$^{-1}$) that is considered in detail in
Figures 4, 5, 8, 10, and 11.}
\end{figure}

The preponderance of evidence for anisotropy is also
illustrated in Figure 7, which shows the probability that each
measurement could be explained by an isotropic O$^{5+}$ velocity
distribution.
In other words, Figure 7 gives the value of $P_{a}(1)$ for
each probability cube.
Taken together, a significant majority of the values (about 78\%
of the total number) fall below the fiducial one-sigma value of
$P_{1\sigma}$, indicating that isotropy should not be
considered a ``baseline'' assumption.
Below about $r = 2 \, R_{\odot}$, a few of the measurements
correspond to large probabilities that an isotropic distribution
can explain the observations.
Note from Figure 6{\em{c}}, though, that the most-probable
anisotropy ratios for these measurements tend to be greater
than 1, but some of the error bars extend down past
$T_{i \perp}/T_{i \parallel}=1$.
However, between 2.1 and 2.7 $R_{\odot}$ the probability of
isotropy is very small for all of the observed data points.
Above 3 $R_{\odot}$, some of the values of $P_{a}(1)$ become large
again, but we believe this may be due to the relatively high
observational uncertainties on the \ion{O}{6} intensities
and line widths at these large heights (see below).

\begin{figure}
\epsscale{1.13}
\plotone{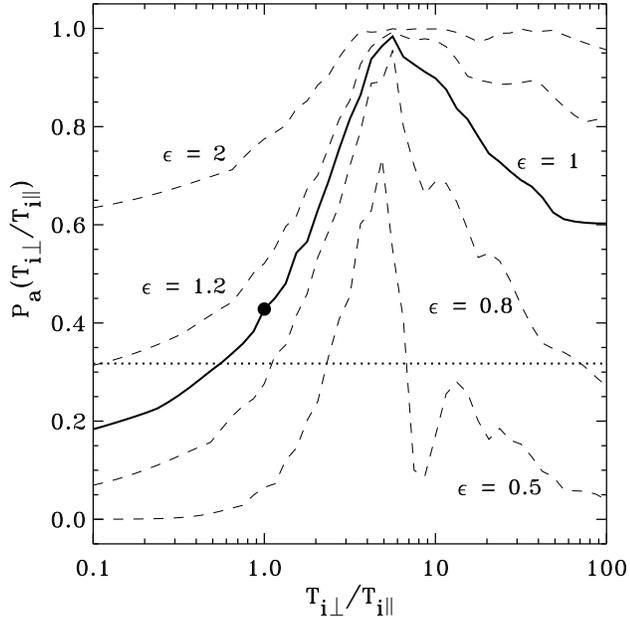}
\caption{Reduced probability $P_{a}$ versus ion anisotropy
ratio $T_{i \perp}/T_{i \parallel}$ for one specific comparison
between a UVCS observation at $r = 3.07 \, R_{\odot}$ and the
model R data cube (see also Figs.\  4, 10, and 11).
The probability computed with the actual observational
uncertainties ({\em thick solid line}) is compared with trial
curves computed with a range of constant factors $\epsilon$
multiplying $\delta V_{1/e}$ and $\delta {\cal R}$
({\em dashed lines}).
Also shown is the threshold level $P_{1\sigma}$
({\em dotted line}) and the probability of isotropy for the
standard $\epsilon=1$ case that is also shown in Figure 7
({\em filled circle}).}
\end{figure}

To better understand the impact of observational uncertainties
on the probability of isotropy, Figure 8 shows the full
$P_{a} (T_{i \perp}/T_{i \parallel})$ curves for one specific
measurement at $r = 3.07 \, R_{\odot}$ (i.e., the same
measurement used in Fig.\  5).
The multiple curves were constructed by multiplying the
known observational uncertainties $\delta V_{1/e}$ and
$\delta {\cal R}$ by arbitrary factors $\epsilon$.
Generally, larger uncertainties lead to lower $\chi^{2}$ values
when comparing the observed and modeled line shapes, and thus to
larger probabilities of agreement between the observed and
modeled profiles.
Interestingly, though, the anisotropy ratio 
$T_{i \perp}/T_{i \parallel}$ at which the maximum probability
occurs remains roughly constant when $\epsilon$ is varied
between 0.5 and 2.
Thus, if future observations above 3 $R_{\odot}$ were to obtain
the same general range of values for $V_{1/e}$ and ${\cal R}$
but with lower uncertainties, it could provide stronger evidence
for ion anisotropy up at these heights.

Although Figure 6{\em{c}} does not seem to indicate a substantial
difference between models R and C, it is useful to compare these
models in some additional detail.
For all data points, the mean ratio of model C to model R
anisotropy ratios was 1.199, but the large standard deviation
(76\%) shows that the models are often quite different from
one another.
Taking only the heights above 2.2 $R_{\odot}$, the mean ratio of
model C to model R anisotropy ratios increases to 1.514,
indicating that {\em on average} model C generates larger
anisotropies than model R over the height range where anisotropies
appear to be required.

\begin{figure}
\epsscale{1.13}
\plotone{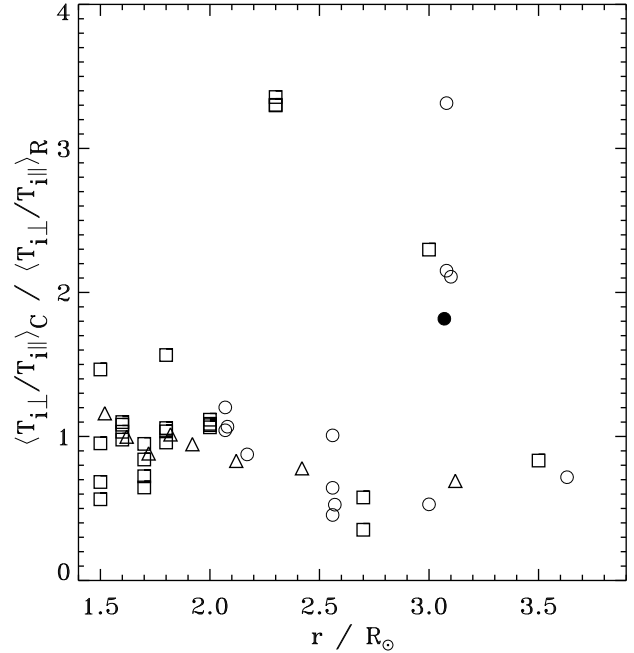}
\caption{Ratios of model C to model R values for the weighted mean
anisotropy ratios $\langle T_{i \perp} / T_{i \parallel} \rangle$
shown as a function of height and using the same data symbols
as Figures 1, 6, and 7.}
\end{figure}

Figure 9 shows the ratio of model C to model R anisotropy ratios
as a function of height.
Below $r \approx 2.2 \, R_{\odot}$ the two models produce
roughly the same result for the anisotropy ratio.
Above that height, the solutions split into two groups: one
where model C produces a substantially larger ratio (2--3
times that of model R), and one where model C produces
a comparable or slightly smaller ratio than model R.
Note that the height range of 3.0--3.1 $R_{\odot}$---over which
model R predicted a rise in the probability of isotropy (see
Fig.\  7)---strongly favors larger anisotropies for model C.

\subsection{Varying the Electron Density, Electron
Temperature, and Geometry}

One of the main motivations for this paper was to explore why
the results of Raouafi \& Solanki (2004, 2006) were so different
from earlier results (e.g., Cranmer et al.\  1999) regarding
the O$^{5+}$ anisotropy ratio.
In this subsection, we study the differences between model R
and model C in more detail by focusing on the shapes of the
reduced probability distributions for one representative data point.
As in Figures 4, 5, and 8, we used the probabilities generated by
comparing the UVCS/{\em{SOHO}} measurement from 1997 January 5
($r = 3.07 \, R_{\odot}$, $V_{1/e} = 690$ km s$^{-1}$) with
data cubes constructed with various assumptions.
This data point is denoted by a filled circle in Figure 9,
and it is clear that this point is representative of the majority
of the data points (5 out of 7) at $r \approx 3 \, R_{\odot}$.

Figure 10 shows a range of reduced probability curves
$P_{a} (T_{i \perp}/T_{i \parallel})$ that were computed from
data cubes constructed with various combinations of the model R
and model C parameters.
The three-letter names for the models denote the individual
choices for $n_e$, flux-tube geometry, and $T_e$ (in that order).
The ``pure'' model R and model C cases are thus called RRR
and CCC.

\begin{figure}
\epsscale{1.13}
\plotone{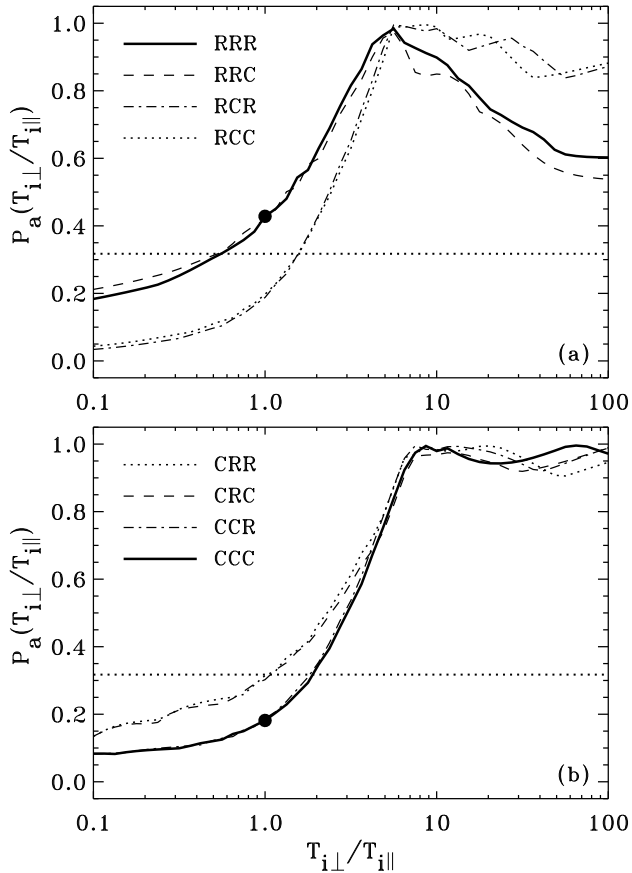}
\caption{Reduced probability $P_{a}$ versus ion anisotropy
ratio $T_{i \perp}/T_{i \parallel}$ for the same data
comparison as in Figures 4 and 8, but for various
combinations of the model R and model C parameters
(see above for line styles).
The order of the three-letter designations is
$\{ n_{e}, f(r), T_{e} \}$.
Panel {\em{(a)}} thus shows all models computed with the
Doyle et al.\  (1999) $n_e$ and panel {\em{(b)}} shows all
models computed with the Cranmer et al.\  (1999) $n_e$.
Also shown is the threshold level $P_{1\sigma}$
({\em horizontal dotted lines}) and the probability of isotropy
for the main RRR and CCC cases ({\em filled circles}).}
\end{figure}

Before examining the impact of the individual parameters on
the reduced probability curves, we note that the model CCC
curve in Figure 10{\em{b}} mirrors almost exactly the results
of Kohl et al.\  (1998) and Cranmer et al.\  (1999) at
$r \approx 3 \, R_{\odot}$:
the most likely O$^{5+}$ anisotropy ratio ranges between 10
and 100, and an isotropic distribution is highly improbable.
Model CCC exhibits a most probable ion outflow speed
$\langle u_{i \parallel} \rangle = 508$ km s$^{-1}$, which is only
marginally smaller than the model RRR value of 521 km s$^{-1}$.
Model CCC has an optimal solution for $\langle w_{i \perp} \rangle$,
though, of 541 km s$^{-1}$, which is 23\% larger than the
corresponding value of 440 km s$^{-1}$ for model RRR (i.e., a
51\% higher value of $\langle T_{i \perp} \rangle$ for model CCC).
Model CCC tended to produce more line broadening via ``thermal''
motions near the plane of the sky, and model RRR tended to produce
more line broadening via bulk outflow projected along the LOS.

The other curves shown in Figure 10 explore which of the three
varied parameters were most responsible for the differences
between models RRR and CCC.
We see immediately that the choice of electron temperature $T_e$,
which in our models impacts only the collision rate $q_{12}$,
is relatively unimportant.
The 8 curves can thus be separated into 4 pairs, each of which
has the same choice for $n_e$ and flux-tube geometry (i.e., RRX,
RCX, CRX, and CCX, where `X' denotes either option for $T_e$).
The overall insensitivity to electron temperature is evident
from the fact that the two curves in each pair are virtually
indistinguishable from one another.

Figure 10 shows that the unique features of the RRX models
(i.e., a higher probability of isotropy and a strong peak
at $T_{i \perp}/T_{i \parallel} < 10$) are only present when
{\em both} the electron density and flux-tube geometry are
treated using model R.
The models with only one of these two parameters treated
using model R (i.e., RCX and CRX) appear more similar to
the CCX models.
At large values of the anisotropy ratio, both the RCX and
CRX models are virtually identical to the CCX models.
At low values of the anisotropy ratio, the CRX model is roughly
intermediate between the CCX and RRX models.
Generally, though, the {\em combination} of the model R assumptions
for electron density (e.g., Doyle et al.\  1999) and flux-tube
geometry (e.g., Banaszkiewicz et al.\  1998) are needed to
produce broad enough profiles via outflow speed projection
along the LOS to explain the observations {\em without} the need
for extreme temperature anisotropies.
Specifically, this enhanced LOS projection effect arises for two
coupled reasons.
\begin{enumerate}
\item
As seen in Figure 2{\em{c}}, the Banaszkiewicz et al.\  (1998)
flux tubes are tilted to a greater degree away from the radial
direction than the Cranmer et al.\  (1999) flux tubes.
Because of these larger values of $\delta$, a larger fraction
of the outflow speed $u_{i \parallel}$ is projected into the
LOS direction (when $|x| > 0$) for model R.
\item
Figure 3 shows that the Doyle et al.\  (1999) electron density
does not drop as rapidly with increasing height (between
about 3 and 10 $R_{\odot}$) as nearly all of the other plotted
$n_e$ functions.
Thus, for observation heights at about 3 $R_{\odot}$, the
Doyle et al.\  (model R) electron density provides a relative
enhancement for points along the foreground and background
($|x| > 0$) in comparison to the plane of the sky ($x=0$).
\end{enumerate}

Note also from Figure 3 that the electron density used for model C
is about 10\% to 30\% larger than that used for model R at the
heights of interest ($r \approx 3$--4 $R_{\odot}$).
A higher value of $n_e$ is expected to result in emission lines
that are dominated more by the collisional component of the
emissivity, which scales as $n_{e}^2$ (eq.~[\ref{eq:jcoll}]), with
a correspondingly weaker contribution from the radiative component,
which scales linearly with $n_e$ (eq.~[\ref{eq:jres}]).
Because of the different density dependences, the collisional
component is not extended as far along the LOS as the radiative
component.
Thus, models with higher densities would be expected to behave more
like model C (with emission dominated by the plane of the sky),
and models with lower densities would be expected to behave more
like model R (with emission extended over a larger swath of the LOS).

To explore the effects of varying the electron density, we repeated
the model R data cube analysis (for the fiducial height shown in
Figs.\  8 and 10) with $n_{e}(r)$ multiplied by constant factors.
Figure 11 shows the resulting reduced probability curves as a function
of the O$^{5+}$ temperature anisotropy.
A model with half of the Doyle et al.\  (1999) electron density
has a lower preferred value of $T_{i \perp}/T_{i \parallel}$
and a much higher probability of isotropy than the standard model R.
A model with double the Doyle et al.\  (1999) electron density
resembles model C in that there is a high preferred range of
$T_{i \perp}/T_{i \parallel}$ and a low probability of isotropy.
Despite the large change in appearance of the $P_{a}$ curves as
shown in Figure 11, the preferred values of the outflow speed and
perpendicular kinetic temperature do not vary by very much as
$n_e$ is varied up and down by a factor of two:
$\langle u_{i \parallel} \rangle$ changes by only about $\pm 8$\%
(increasing as $n_e$ decreases), and $\langle w_{i \perp} \rangle$
changes by only about $\pm 10$\% (increasing as $n_e$ increases).
These determinations appear to be relatively insensitive to the
choices for $n_e$ and flux-tube geometry.

\begin{figure}
\epsscale{1.13}
\plotone{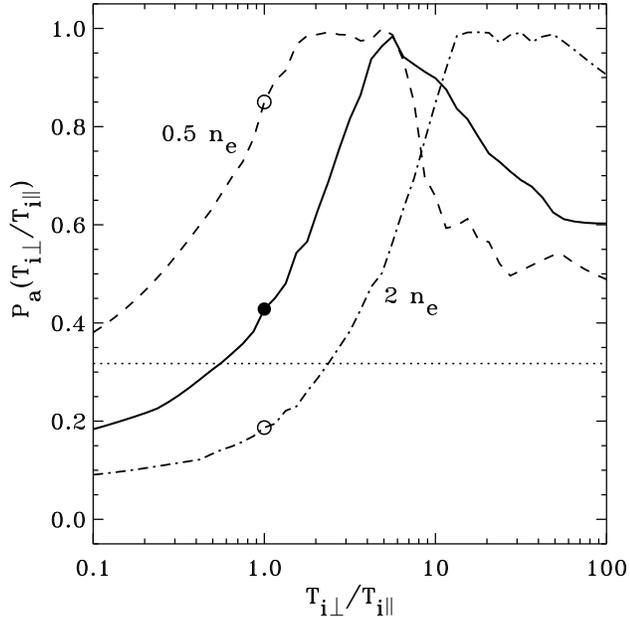}
\caption{Reduced probability $P_{a}$ versus ion anisotropy
ratio $T_{i \perp}/T_{i \parallel}$ for the same data
comparison as in Figures 4, 5, 8, and 10, but for a range of
constant multipliers to the Doyle et al.\  (1999)
electron density.
The basic ``model R'' $n_e$ ({\em solid line}) is compared
to a model with half ({\em dashed line}) and double
({\em dot-dashed line}) this electron density function.
Also shown is the threshold level $P_{1\sigma}$
({\em dotted line}) and the probabilities of isotropy for
the three curves ({\em filled and open circles}).}
\end{figure}

The ratio of collisional emissivity to the total line emission
changes dramatically for the models shown in Figure 11.
For model RRR, the optimal model in the data cube exhibited a
collisional fraction of 93.7\% for the \ion{O}{6} $\lambda$1032
line and a fraction of 44.6\% for the \ion{O}{6} $\lambda$1037
line (the latter being ``Doppler pumped'').
The model with half of the model R density had lower collisional
fractions for the $\lambda\lambda$1032, 1037 lines of
88.2\% and 28.7\%, respectively.
The model with double the model R density had higher collisional
fractions of 96.8\% and 61.7\%.

It is important to note, however, that the differences in
collisionality for the models shown in Figure 10 are not as
drastic as those shown in Figure 11.
Model CCC exhibited collisional fractions for the
$\lambda\lambda$1032, 1037 lines of 90.7\% and 47.0\%.
These values are only a few percentage points different from the
model RRR fractions.
The other intermediate models have values that cluster between
those of models RRR and CCC.
The larger value of $n_e$ in the plane of the sky for model C
is compensated---to some degree---by the slower decrease in
$n_e$ along the LOS for model R.
Thus, despite the superficial resemblance between the model CCC
curve in Figure 10 and the ``double $n_e$'' curve in Figure 11,
one cannot invoke a varying amount of collisionality to
explain the differences between models R and C.
The LOS extension effects discussed above are more subtle
than simply varying $n_e$ by a constant amount.

Another way we explored the dependence of the reduced probabilities
on electron density was to produce a set of three other models
with alternate functional forms for $n_{e}(r)$, but the same
flux-tube geometry and $T_e$ as used in model R.
These models utilized the mean electron density curves from
Guhathakurta \& Holzer (1994), Fisher \& Guhathakurta (1995),
and Guhathakurta et al.\  (1999) (see also Fig.\  3), and
the \ion{O}{6} data cubes were created only at the fiducial
height of 3.09 $R_{\odot}$.\footnote{%
We also created a data cube for the hydrostatic equation (1) of
Doyle et al.\  (1999), but this model exhibited an unusually
strong extension along the LOS.  There was a substantial
contribution to the \ion{O}{6} emissivity even at the LOS
integration limits of $x = \pm 15 \, R_{\odot}$, which actually
led to an extremely {\em low} probability of isotropy.
However, we discarded this model because the shallow $n_e$ at
large heights is clearly unphysical.}
The reduced probabilities $P_a$ for these models all fell within
the general range of variation illustrated in Figure 10 and are
not plotted.
However, the construction of these models increased the number
of data cubes with ``model R-like'' flux-tube and $T_e$ parameters
to seven: i.e., these three new ones, the three models shown
in Figure 11, and model CRR (with a model C electron density).
We performed a regression analysis on the seven values of the
probability of isotropy $P_{a}(1)$ and the weighted mean
anisotropy $\langle T_{i \perp} / T_{i \parallel} \rangle$
to find the optimal functional dependence on two ``independent''
variables that characterize the electron density:
\begin{equation}
  n_{3} \, \equiv \,
  \frac{n_{e}(3 \, R_{\odot})}{10^{6} \, \mbox{cm}^{-3}} \,\, ,
  \,\,\,\,\,\,\,
  n_{6} \, \equiv \,
  \frac{n_{e}(6 \, R_{\odot})}{10^{6} \, \mbox{cm}^{-3}}
\end{equation}
where the arbitrary normalizations serve only to keep the combined
quantities (discussed below) of order unity.
The quantity $n_3$ characterizes the electron density in the plane
of the sky of the observation, and the ratio $n_{3}/n_{6}$
characterizes the large-scale density gradient and thus the relative
enhancement of foreground and background regions along the LOS.
From the discussion above, we expect that larger values of both
$n_3$ and $n_{3}/n_{6}$ should result in lower probabilities of
isotropy and higher most-probable values of the anisotropy ratio.
Indeed, the regression analysis found that the modeled values
of these quantities exhibited the lowest combined $\chi^{2}$
spread for a single independent variable that scales as
$n_{3}^{1.83}/n_{6}$ (close to the product of $n_3$ and
$n_{3}/n_{6}$).
Figure 12 shows these values as well as the best-fitting quadratic
functions to $P_{a}(1)$ and
$\langle T_{i \perp} / T_{i \parallel} \rangle$.
The combined dependence on both $n_e$ and its radial gradient
appears to be a key factor in determining the relative probabilities
of isotropy and strong anisotropy.

\begin{figure}
\epsscale{1.10}
\plotone{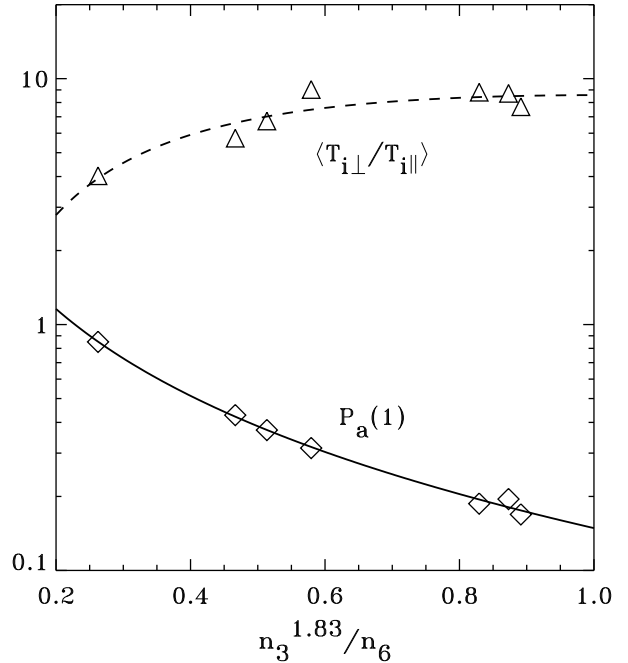}
\caption{Reduced probabilities of isotropy ({\em diamonds})
and weighted mean anisotropy ratios ({\em triangles}) for
models having a range of $n_e$ values and identical flux-tube
and $T_e$ properties (see text for details).
All data-cube comparisons were computed for the same fiducial
data point illustrated in Figures 4, 5, 8, 10, and 11.
Curves denote the best fitting quadratic relations as a
function of the optimized density quantity $n_{3}^{1.83}/n_{6}$.}
\end{figure}

Finally, we must evaluate which sets of choices for the
electron density and the flux-tube geometry are most
consistent with observed polar coronal holes.
Figure 3 does seem to indicate that most measured $n_e$
curves (as well as one example theoretical result for $n_e$)
behave more like the ``model C'' (Cranmer et al.\  1999) case
than the ``model R'' (Doyle et al.\  1999) case.
Between the heights of about 3 and 10 $R_{\odot}$, the majority
of the curves in Figure 3 exhibit a {\em steeper} radial
decrease than the Doyle et al.\  (1999) model.
Thus, the CCX or CRX models shown in Figure 10{\em{b}} appear to
be more consistent with observations than the RRX or RCX models
in Figure 10{\em{a}}.
This then implies that substantial O$^{5+}$ anisotropy
($T_{i \perp} / T_{i \parallel} \gtrsim 10$) is also preferred
at large heights.
The optimal choice of the flux-tube geometry is less certain.
Ideally, observations of the nonradial shapes of polar plumes
should be able to constrain the magnetic field geometry
(see, e.g., Wang et al.\  2007; Pasachoff et al.\  2007),
but it is unclear whether existing plume observations would be
able to distinguish the subtle differences between, e.g.,
Figures 2{\em{a}} and 2{\em{b}}.
In any case, the geometry does not seem to be as major an issue
as the electron density, since the variance between the four
curves in Figure 10{\em{b}} is not large.

\subsection{Oxygen Ion Number Density}

By comparing the observed and modeled total intensities of
the \ion{O}{6} $\lambda$1032 line, it is possible to derive
firm limits on the combined elemental abundance and ionization
fraction of O$^{5+}$.
The ion concentration is useful both as a tracer of fast and slow
solar wind streams (e.g., Zurbuchen et al.\  2002) and as a
possible diagnostic of the amount of preferential heating
deposited in coronal holes (Lie-Svendsen \& Esser 2005).
A first attempt at determining the O$^{5+}$ number density from
UVCS data was made by Cranmer et al.\  (1999), but the
``data cube search'' technique developed in this paper allows
a much more definitive set of measurements to be made.

The numerical code that computes the \ion{O}{6} line emission
used an arbitrary constant value for the ratio
$f_{0} = n_{{\rm O}^{5+}} / n_{e}$ of $4.959 \times 10^{-6}$,
which was derived from the oxygen abundance of
Anders \& Grevesse (1989) and the measured O$^{5+}$ ionization
fraction of Wimmer-Schweingruber et al.\  (1998).
This is merely a fiducial value that does not affect the final
determination of this ratio for a given UVCS observation.
When comparing the results from an empirical model data cube
with a specific observation, the probability values
$P(u_{i \parallel}, T_{i \perp}, T_{i \perp}/T_{i \parallel})$
are used to construct a weighted mean of the modeled
\ion{O}{6} $\lambda$1032 total intensity using
equation (\ref{eq:Pmom}),
as well as lower and upper limits using the full range of models
with probabilities that exceed $P_{1\sigma}$.
These values are converted into ``observed'' ion concentration
ratios $f_{\rm obs}$ by assuming that the ratio
$f_{\rm obs} / f_{0}$ is equal to the ratio of
the observed to the modeled values of $I_{\rm tot}$.
By using the modeled weighted mean, lower limit, and upper limit
of $I_{\rm tot}$ we obtain the weighted mean, upper limit,
and lower limit of $f_{\rm obs}$.
(Note that the lower limit of $I_{\rm tot}$ gives the upper limit
of $f_{\rm obs}$ and vice versa.)
Finally, $f_{\rm obs}$ is converted into the ratio of O$^{5+}$ to
total hydrogen number density ($n_{{\rm O}^{5+}} / n_{\rm H}$)
by multiplying by a factor of 1.1 (assuming a helium to hydrogen
number density ratio of 5\%).

Figure 13 shows the resulting ion concentration ratios as a
function of height for the full range of model C data points.
There is a hint of systematic radial increase at low heights.
A similar radial increase would be predicted for ions that flow
substantially faster than protons in the corona (by a factor of
two) and then flow only $\sim$10\% faster than protons at 1 AU.
Above 2 $R_{\odot}$, though, Figure 13 does not show any
definitive radial trend.
Taking account of the uncertainty limits, the data appear
consistent with the O$^{5+}$ ionization fraction being more or
less ``frozen in'' (see, e.g., Hundhausen et al.\  1968;
Owocki et al.\  1983; Ko et al.\  1997).
A linear least squares fit (using the logarithm of
$n_{{\rm O}^{5+}} / n_{\rm H}$ as the ordinate) is also shown,
but the relatively high uncertainties at large heights preclude
any reliable interpretation of the slope.

\begin{figure}
\epsscale{1.10}
\plotone{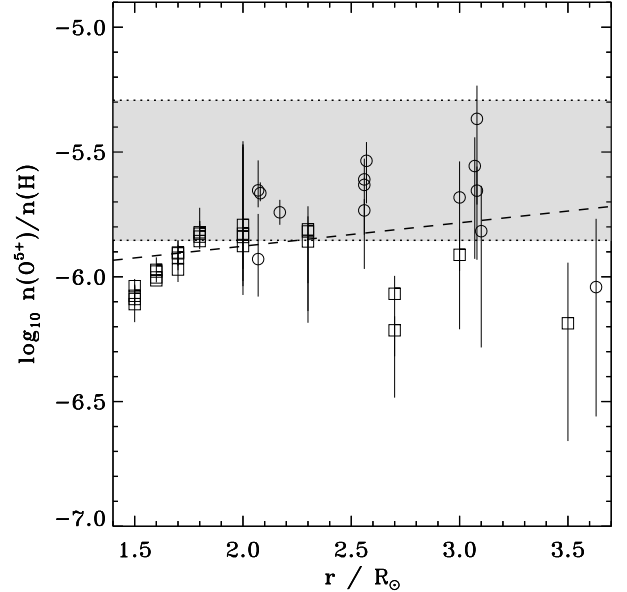}
\caption{O$^{5+}$ ion number density ratio (with respect to hydrogen)
as a function of heliocentric distance for model C.
Symbols show the weighted means of the reduced probability
distributions, with styles the same as in Figure 1.
Vertical bars show the full range of parameter space with
reduced probabilities greater than $P_{1\sigma}$.
Also shown is a linear least-squares fit to the data points
({\em dashed line}) and empirical lower and upper limits as
described in the text ({\em gray region bounded by dotted lines}).}
\end{figure}

Figure 13 also shows a range of values that would have been
expected from prior studies of both the oxygen abundance and
the O$^{5+}$ ionization fraction.
The abundance ratio ($n_{\rm O} / n_{\rm H}$) ranges from
a relatively recent historical high of
$8.5 \times 10^{-4}$ (Anders \& Grevesse 1989)
to the more recent---and somewhat controversial---low of
$4.6 \times 10^{-4}$ (Asplund et al.\  2004, 2005;
Grevesse et al.\  2007).
The ionization fraction ($n_{\rm O^{5+}} / n_{\rm O}$) was
measured in~situ by the SWICS instrument on {\em Ulysses}
(Wimmer-Schweingruber et al.\  1998) to be about 0.0058 in the
fast solar wind.
Models that include the freezing in of heavy ions
have produced values for this ratio from 0.0035
(Esser \& Leer 1990) to about 0.005 (Chen et al.\  2003).
Although collisional ionization equilibrium is not expected to
hold in the extended corona (for polar coronal holes), it is
interesting to note that for $T_{e} = 10^{6}$ K, both
Arnaud \& Rothenflug (1985) and Mazzotta et al.\  (1998)
give a ratio of about 0.0045.
This is similar to the above values, but it varies up
and down by about a factor of 50\% as $T_e$ is decreased or
increased by only $\pm 30$\%.
We thus take tentative lower and upper limits for
$n_{\rm O^{5+}} / n_{\rm O}$ of 0.003 and 0.006.
The horizontal lines shown in Figure 13 were computed from
the products of the two lower limits and the two upper limits
given above for
\begin{equation}
  \frac{n_{\rm O^{5+}}}{n_{\rm H}} \, = \,
  \left( \frac{n_{\rm O}}{n_{\rm H}} \right)
  \left( \frac{n_{\rm O^{5+}}}{n_{\rm O}}  \right) \,\, .
\end{equation}

The model C data points shown in Figure 13 have a mean value
of $n_{{\rm O}^{5+}} / n_{\rm H} = 1.52 \times 10^{-6}$ (taking
a simple average) or $1.39 \times 10^{-6}$ (taking the average
of the logarithms).
Performing the same analysis using model R yielded values that
were larger by about 5\% (on average for all data points) to 20\%
(specifically for points at heights above $\sim$3 $R_{\odot}$).
The standard deviations of both sets of data points gave lower
and upper bounds of approximately $8 \times 10^{-7}$ and
$2.4 \times 10^{-6}$ that encompass the $\pm 1 \sigma$ range.
The prior studies of oxygen abundance and O$^{5+}$ ionization
discussed above give somewhat higher values, which extend from
$1.4 \times 10^{-6}$ to $5.1 \times 10^{-6}$.
Thus, both the model C and model R data points are in reasonably
good agreement with the {\em lower limit} of the expected range,
which gives some support for the recent low oxygen abundances of
Asplund et al.\  (2004, 2005).

\section{Discussion and Conclusions}

The {\em SOHO} mission (Domingo et al.\  1995) has made significant
progress toward identifying and characterizing the processes that
heat the corona and accelerate the solar wind (see also
Fleck \& \v{S}vestka 1997; Domingo 2002; Fleck 2004, 2005).
The results from the UVCS instrument regarding preferential
heating and acceleration of heavy ions (i.e., O$^{5+}$) have
contributed in a major way to these advances in understanding
over the past decade, and it is important to verify and confirm
the key features of these results.
Thus, this paper has analyzed an expanded set of UVCS
data from polar coronal holes at solar minimum with the goal of
ascertaining whether ion temperature anisotropies are
definitively present (as claimed by Kohl et al.\  1997, 1998;
Li et al.\  1998; Cranmer et al.\  1999; Antonucci et al.\  2000;
Zangrilli et al.\  2002; Antonucci 2006; Telloni et al.\  2007)
or whether one can explain the observations without such
anisotropies (as claimed by Raouafi \& Solanki 2004, 2006;
Raouafi et al.\  2007).
These cases were exemplified by two sets of empirical models:
one (model R) that was designed to replicate many of the
conditions assumed by Raouafi \& Solanki (2004, 2006), and
one (model C) that used the same conditions as
Cranmer et al.\  (1999).

The main conclusion of this paper is that there remains strong
evidence in favor of both preferential O$^{5+}$ heating and
acceleration and significant O$^{5+}$ ion temperature anisotropy
(in the sense $T_{i \perp} > T_{i \parallel}$) above
$r \approx 2.1 \, R_{\odot}$ in coronal holes.
More detailed conclusions, linked to the sections of the paper
in which they were first discussed, are summarized as follows.
\begin{enumerate}
\item
It is important to search the full range of possible O$^{5+}$
ion properties and not make arbitrary assumptions about, e.g.,
the ion outflow speed or the ion temperature.
It is clear from Figure 5{\em{b}} that if the comparison with
observations is restricted to certain choices for the ion
parameters, the resulting conclusions about the ion temperature
anisotropy can be potentially misleading. ({\S}~4.1)
\item
The derived ion outflow speeds $u_{i \parallel}$ and
perpendicular kinetic temperatures $T_{i \perp}$ exhibit values
similar to those reported by Kohl et al.\  (1998) and
Cranmer et al.\  (1999), independent of the choices of electron
density and flux-tube geometry.
There is significant evidence for preferential ion heating
and ion acceleration with respect to protons, although
the radial rate of increase of $T_{i \perp}$ may be
slightly lower than that given by Cranmer et al.\  (1999).
The large values of $T_{i \perp}$ appear to be due to true
``thermal'' motions and not unresolved wave motions. ({\S}~4.2)
\item
For heights above about 2.1 $R_{\odot}$, the models in
this paper yielded higher probabilities of agreement with the
UVCS observations for {\em anisotropic velocity distributions}
than for isotropic distributions.
The UVCS observations between the radii of 2.1 and 2.7
$R_{\odot}$ were found to have probabilities of isotropy below
about 10\% (see Fig.\  7), no matter what was assumed for the
coronal electron density or flux-tube expansion (i.e., for
either model R or model C).
Even when using coronal properties that seemed to maximize the
probability of isotropy (e.g., model R), 78\% of the UVCS data
points exhibited probabilities of isotropy below our threshold
one-sigma value of $\sim$32\%. ({\S}~4.3)
\item
The UVCS data at heights at and above 3 $R_{\odot}$
can be used to put limits on the likelihood of strong
O$^{5+}$ temperature anisotropies.
A key factor in discriminating between empirical models that
either require or do not require a substantial anisotropy is the
degree of {\em extension along the line of sight (LOS)} of the
emissivity.
This extension is driven strongly by the rate of radial decrease
in the electron density.
The relatively shallow slope of $n_{e}(r)$ used in model R
(from eq.~[\ref{eq:neD99}]) appears to be an ``outlier''
when compared to other observational and theoretical
determinations of the electron density profile in coronal
holes (see Fig.\  3).
Most other $n_{e}(r)$ curves exhibit a steeper radial decrease
and thus a lesser degree of LOS extension for the \ion{O}{6}
emissivities.
Our model C, which utilized the empirical model parameters
derived by Cranmer et al.\  (1999), had a representative
``steep'' electron density profile and thus required a
substantial O$^{5+}$ temperature anisotropy to explain the
UVCS observations above $r \approx 3 \, R_{\odot}$.  ({\S}~4.4)
\item
Models that exhibit enough LOS extension to reproduce the observed
UVCS line profiles and intensities {\em without a temperature
anisotropy} appear to require both (1) an electron density that
decreases shallowly with increasing height, and (2) a highly
superradial flux-tube geometry that projects a large fraction
of the outflow speed vector into the LOS.
Our model R, designed to be similar to the models used by
Raouafi \& Solanki (2004, 2006), exhibited both of these
conditions and thus had higher probabilities of an isotropic
velocity distribution at heights above
$r \approx 3 \, R_{\odot}$. ({\S}~4.4)
\item
At the largest heights ($r \gtrsim 3 \, R_{\odot}$), the
uncertainties in the existing UVCS measurements make difficult
a firm determination of the anisotropy ratio.
The analysis technique developed in this paper takes full
account of these observational uncertainties.
Future observations with smaller observational uncertainties
(see Fig.\  8) should yield correspondingly ``sharper''
probability distributions for the anisotropy ratio and thus
better determinations of this quantity. ({\S}~4.3)
\item
Total intensities of the \ion{O}{6} $\lambda\lambda$1032, 1037
lines constrain the ion concentration ratio
$n_{{\rm O}^{5+}} / n_{\rm H}$ to be approximately
$1.5 \times 10^{-6}$, with at least a factor of two range of
uncertainty.
If the freezing in of O$^{5+}$ ions is considered to be relatively
well understood, then this value implies a relatively low oxygen
abundance in agreement with the recent downward revision of
Asplund et al.\  (2004, 2005).  ({\S}~4.5)
\end{enumerate}

Because of existing observational uncertainties in the electron
density, flux tube geometry, and \ion{O}{6} line parameters
such as $V_{1/e}$ (the line width) and ${\cal R}$
(the $\lambda$1032 to $\lambda$1037 intensity ratio),
we cannot yet give ``preferred'' values for the O$^{5+}$ anisotropy
ratio $T_{i \perp} / T_{i \parallel}$ as a detailed function
of height.
Below $r \approx 2 \, R_{\odot}$, the observations are
consistent with an isotropic velocity distribution.
Between 2.2 and 2.7 $R_{\odot}$, the most probable anisotropy
ratio appears to range between 2 and 10 (see Fig.\  6{\em{c}}).
At heights around $r \approx 3 \, R_{\odot}$ the uncertainties
are large, but there does seem to be evidence that the anisotropy
ratio is likely to exceed 10 (see, e.g., Fig.\  10{\em{b}}).
The ratio must increase between 2 and 3 $R_{\odot}$, but we do not
yet claim to know the exact rate of increase.

New observations are required to make further progress.
For example, as seen in Figure 3, there is still some disagreement
about the radial dependence of electron density in polar
coronal holes.
Measurements of the white-light polarization brightness ($pB$)
at solar minimum need to be made with lower uncertainties in
the absolute radiometric calibration.
Also, care must be taken to exclude time periods when high-latitude
streamers may be contaminating the LOS in order to obtain a true
mean electron density for a polar coronal hole.
Existing measurements of the superradial geometry (as traced by,
e.g., polar plumes) tend to be limited by the fact that the
shapes evident in LOS-integrated images are often assumed to be 
identical to the shapes of flux tubes in the plane of the sky.
Better constraints on the flux-tube geometry could thus be
made by using stereoscopy (Aschwanden 2005),
tomography (e.g., Frazin et al.\  2007),
or other time-resolved rotational techniques (e.g.,
DeForest et al.\  2001) to trace the full three-dimensional
shapes of the plume-filled flux tubes.

Improved ultraviolet spectroscopic measurements would
greatly improve our ability to determine the plasma parameters
in coronal holes.
We anticipate that the UVCS instrument will continue to observe
polar coronal holes through the present solar minimum (2007--2008).
We do not yet know whether the wide spread in line widths seen
a decade ago (which exceeded the observational uncertainties) was
due to a changing filling factor of polar plumes along the LOS or
whether it may be connected to other kinds of time variability
at the coronal base.
An even wider range of variation in coronal hole properties was
observed over the last solar cycle with UVCS (e.g.,
Miralles et al.\  2006).
These observations of how coronal holes evolve in size and
latitude have helped to constrain the realm of possible parameter
space of preferential ion heating and acceleration.

There are also observations that cannot be made with UVCS that
could greatly improve our understanding of ion energization
in the solar wind acceleration region.
For example, rather than having only O$^{5+}$ (and, to a lesser
extent, Mg$^{9+}$; see Kohl et al.\  1999) in coronal holes,
an instrument with greater sensitivity and a wider spectral range
could sample the velocity distributions of dozens of additional
ions with a range of charges and masses.
Obtaining the distribution of derived kinetic temperatures as a
function of the ion charge-to-mass ratio $Z/A$ would put a firm
constraint on the shape of the power spectrum of cyclotron-resonant
fluctuations (e.g., Hollweg 1999; Cranmer 2002b).
A next-generation instrument with greater sensitivity may also
be able to detect subtle departures from Gaussian line shapes
that signal the presence of specific non-Maxwellian distributions
(e.g., Cranmer, 1998, 2001).

The strong heating and acceleration of minor ions, as
documented by UVCS/{\em{SOHO}}, has provided significant insight
into the physics of solar wind acceleration, but the basic
chain of physical processes is still somewhat unclear.
Many theoretical studies have attempted to trace this chain
``backwards'' from the known facts of kinetic ion energization
to the properties of, e.g., ion cyclotron resonant waves that
can provide such energization naturally (see reviews by
Hollweg \& Isenberg 2002; Marsch 2005; Kohl et al.\  2006).
Complementary progress has also been made in constraining the
large-scale properties of the MHD fluctuations that may
eventually cascade down to the microscopic kinetic scales
(e.g., Verdini \& Velli 2007; Cranmer et al.\  2007).
There are still many areas of disconnect, though, between
our understanding of the macroscopic MHD scales and the
microscopic kinetic scales.
Future theoretical work is expected to continue exploring how
the combined state of turbulent fluctuations, wave-particle
interactions, and species-dependent heating and acceleration
can be produced and maintained.

\acknowledgments

The authors would like to thank Adriaan van Ballegooijen,
Mari Paz Miralles, Leonard Strachan, and John Raymond
for valuable discussions.
This work has been supported by the National Aeronautics and Space
Administration (NASA) under grants {NNG\-04\-GE84G},
{NNG\-05\-GG38G}, {NNG\-06\-GI88G}, {NNX\-06\-AG95G}, and
{NNX\-07\-AL72G}
to the Smithsonian Astrophysical Observatory,
by Agenzia Spaz\-i\-ale Italiana, and by the Swiss contribution
to the ESA PRODEX program.

\vspace*{0.20in}
\begin{figure}[h]
\epsscale{1.11}
\figurenum{1}
\plotone{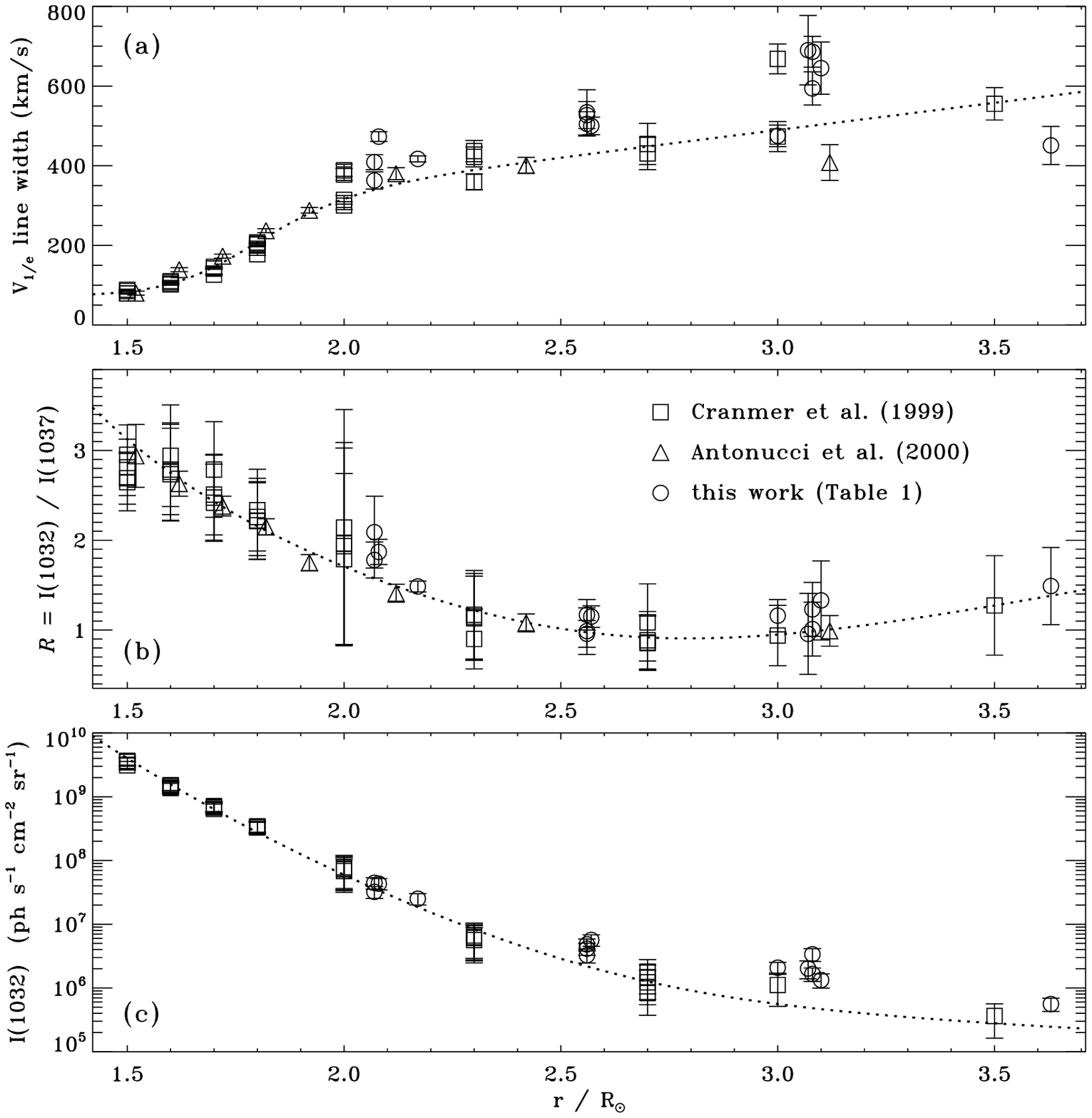}
\caption{Collected UVCS polar coronal hole measurements of
{\em (a)} \ion{O}{6} line widths $V_{1/e}$,
{\em (b)} ratio of \ion{O}{6} $\lambda$1032 to
\ion{O}{6} $\lambda$1037 intensities, and
{\em (c)} \ion{O}{6} $\lambda$1032 line-integrated intensities,
with symbols specifying the sources of the data (see labels
for references).
Error bars denote $\pm 1 \sigma$ observational uncertainties.
Also shown ({\em dotted lines}) are the parameterized fits
given by Cranmer et al.\  (1999).}
\end{figure}

\end{document}